\documentclass[prl,twocolumn, amsmath,amssymb,nofootinbib,superscriptaddress]{revtex4-2}
\usepackage{graphicx,subfigure,amsmath,bm,colordvi,times}
\usepackage{comment}
\usepackage{mathtools}
\usepackage[usenames]{color}
\usepackage{multirow}
\usepackage{epstopdf}
\usepackage[utf8]{inputenc}
\usepackage[english]{babel}
\usepackage{amsfonts}
\usepackage{amssymb}
\usepackage{float}
\usepackage{hyperref}
\usepackage{braket}
\usepackage[normalem]{ulem}

\hypersetup{pdfborder=0 0 0,colorlinks=true,citecolor=blue,linkcolor=blue}

\makeindex

\begin{document}
\title{Signature in sound-mode of the exciton bilayer two-dimensional superfluid transition} 
\author{Giovanni Midei$^\dagger$}
\affiliation{Pitaevskii BEC Center, CNR-INO, Trento, Italy}
\affiliation{COMMIT Group, Department of Physics, University of Antwerp, Groenenborgerlaan 171, 2020 Antwerp, Belgium}
\affiliation{CQM Group, School of Pharmacy, University of Camerino, 62032 Camerino (MC), Italy}

\author{Filippo Pascucci$^\dagger$}
\affiliation{COMMIT Group, Department of Physics, University of Antwerp, Groenenborgerlaan 171, 2020 Antwerp, Belgium}

\author{Milorad V. Milo\v{s}evi\'c}
\affiliation{COMMIT Group, Department of Physics, University of Antwerp, Groenenborgerlaan 171, 2020 Antwerp, Belgium}

\author{Jacques Tempere}
\affiliation{TQC, Department of Physics, University of Antwerp, Universiteitsplein 1, 2610 Antwerp, Belgium}

\author{David Neilson}
\affiliation{COMMIT Group, Department of Physics, University of Antwerp, Groenenborgerlaan 171, 2020 Antwerp, Belgium}

\author{Andrea Perali}
\email{andrea.perali@unicam.it}
\affiliation{CQM Group, School of Pharmacy, University of Camerino, 62032 Camerino (MC), Italy}

\date{\today}

\begin{abstract}
Obtaining definitive evidence of exciton superfluidity in electron--hole bilayers in zero magnetic field remains a major longstanding challenge since the condensate is electrically neutral, making its phase coherence difficult to detect directly.
We show that the Anderson--Bogoliubov sound velocity provides a dynamical signature of exciton superfluidity. Across the BCS--BEC crossover, the velocity is known to discontinuously drop to zero at the Berezinskii--Kosterlitz--Thouless (BKT) transition. The  magnitude of the drop has a strong density dependence. We compute this behavior, with the inclusion of finite-temperature screening, and determine the BKT transition using a renormalization-group approach. We further identify a temperature window which is experimentally accessible, where vortex--antivortex excitations strongly renormalize both the sound velocity and the transition temperature.
\end{abstract}
\maketitle

\def\thefootnote{$\dagger$}
\footnotetext{These authors contributed equally to this work.}

There are growing experimental indications of superfluidity or Bose--Einstein condensation (BEC) in semiconductor electron--hole double layer heterostructures in zero magnetic field~\cite{Burg2018,Wang2019,Gu2022,Ma2021,Nguyen2025}. Electron--hole bilayers are particularly interesting because exciton condensed phases may occur at relatively high temperatures~\cite{Perali2013}, potentially leading to low-dissipation, electrically tunable devices~\cite{Tutuc2004,Su2008,Nandi2012}. Exciton bilayers also provide a versatile platform to explore the BCS--BEC crossover: by tuning the carrier density with external gates, the system can be driven from a strongly coupled BEC regime of tightly bound excitons to a weakly coupled BCS regime of overlapping electron--hole pairs~\cite{Pieri2007, Salasnich2013,LopezRios2018}.

The strong circumstantial indications of the presence of a superfluid or BEC, i.e.\ enhanced interlayer tunneling \cite{Spielman2000,Burg2018,Wang2019}, near perfect Coulomb drag~\cite{Narozhny2016,Li2017,Liu2022,Nguyen2025}, and, recently, spin-valley susceptibility in TMD bilayers~\cite{Qi2026}, are consistent with exciton condensation.  However, none of these directly establishes existence of a superfluid. Definitive confirmation is elusive and a great challenge, particularly since in a neutral exciton superfluid there is no Meissner effect.

We propose employing the Berezinskii--Kosterlitz--Thouless (BKT) transition~\cite{Berezinsky1972,Kosterlitz1973} to develop an experimentally accessible signature for exciton superfluidity. The BKT transition is a topological transition in which bound vortex--antivortex pairs unbind and destroy the quasi-long-range phase coherence in two-dimensional superfluids. Below the BKT transition temperature $T_{\mathrm{BKT}}$, the finite phase stiffness supports a long-wavelength Anderson--Bogoliubov (AB) phase mode~\cite{VanLoon2023}. At $T_{\mathrm{BKT}}$, the Nelson--Kosterlitz universal jump drives the renormalized stiffness to zero~\cite{Nelson1977}, and since the AB sound velocity is proportional to the square root of the stiffness,  there is a discontinuous collapse of the AB velocity at the transition. The collapse of the AB mode can thus provide a fingerprint for the superfluid.

We have developed a finite-temperature path-integral theory for electron--hole bilayers.  This combines temperature-dependent Random Phase Approximation (RPA) screening with a renormalization-group (RG) treatment of the vortex--antivortex fluctuations. With this we  evaluate the BKT transition temperature, the renormalized stiffness, and the AB sound mode. In this way we have a route to identify in a neutral exciton condensate, both the superfluid transition and the vortex-fluctuation regime.

\begin{figure}[t]
    \centering
    \includegraphics[trim=0.0cm 0.0cm 0.0cm 0.0cm, clip=true, width=0.48\textwidth]{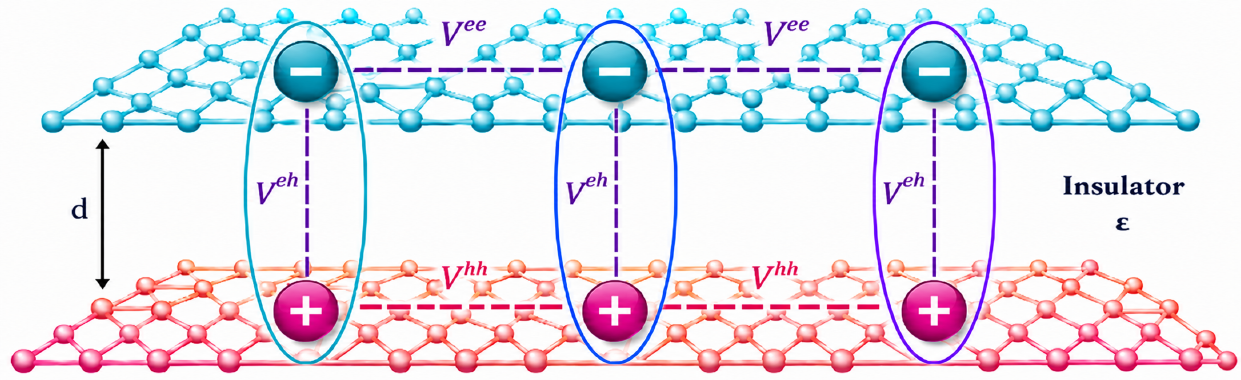}
        \caption{Schematic of the exciton bilayer system. Top layer of electrons, bottom layer of holes. Separated by an insulator of dielectric constant $\varepsilon$ and thickness $d$.}
    \label{Fig.a}
\end{figure}

The electron-hole bilayer system consists of two $n$-doped and $p$-doped conducting parallel layers that are separated by a thin insulating layer  (Fig.\ \ref{Fig.a}).
A particle-hole transformation is used to map the empty electron states in the valence band of the $p$-doped layer to a conduction band populated by positively charged holes. 
The system is then modeled as two oppositely charged species interacting through Coulomb interactions.

The speed of sound is given by the derivative of the AB phase collective mode in the long-wave regime~\cite{Anderson1958, Klimin2019}, and so we start by generalizing the phase-only action for long-range Coulomb interactions. We set  $k_{\mathrm{B}}=\hbar=1$ throughout this work.   
We focus on the low-density regime where the intralayer repulsion is negligible~\cite{Pascucci2024}.
For a superfluid exciton bilayer, the finite-temperature path-integral action is $S=S_0+S_{\mathrm{int}}$ where 
\begin{align}
    S_0 \!=\! & 
    \sum_{\lambda, \sigma}\!
    \int \color{black}
    \mathrm{d}\tau \!\int \mathrm{d}\bm{x}\
    \Psi^{\dagger}_{\lambda,\sigma,\tau}(\bm{x})\!
    \left(\partial_{\tau}+T_{\lambda}-\mu_{\lambda}\right)
    \Psi_{\lambda,\sigma,\tau}(\bm{x}), \label{S0}\\
    S_{\mathrm{int}}\!=\!& 
  \int\limits \mathrm{d}\tau
\int \mathrm{d}\bm{x} \int \mathrm{d}\bm{y}
\Bigg[
   \sum_{\sigma\neq\sigma'}\color{black}
    \Big(
        \Delta^{*}_{\tau}(\bm{x},\bm{y})
        \Psi_{e,\sigma,\tau}(\bm{x})\Psi_{h,\sigma',\tau}(\bm{y}) \nonumber \\
    & + \Delta_{\tau}(\bm{x},\bm{y})
        \Psi^{\dagger}_{e,\sigma,\tau}(\bm{x})
        \Psi^{\dagger}_{h,\sigma',\tau}(\bm{y})
    \Big)
    + \frac{|\Delta_{\tau}(\bm{x},\bm{y})|^{2}}{V_{eh}(\bm{x},\bm{y})}
    \label{SI}
\Bigg].
\end{align}
%
%

$S_0$ is the non-interacting term and $S_{\mathrm{int}}$ the attractive interaction term in the pairing (Bogoliubov) channel \cite{Stratonovich1957}. The fermionic fields $\Psi_{\lambda,\sigma,\tau}(\bm{x})$ are functions of the inlayer position $\bm{x}$ and the imaginary time $\tau$, and are labeled by the spin-valley index
$\sigma=(v,s_z)$ and $\lambda=e,h$ for the electron and hole layers. In Eq.\ \eqref{S0} the kinetic operator $T_\lambda=-\nabla^2/(2m^*)$ and the chemical potentials are $\mu_{\lambda}$. We consider only equal electron and hole effective masses $m^*=m_e^*=m_h^*$ and equal densities $n=n_e=n_h$. In Eq.\ \eqref{SI}, the bare interlayer Coulomb attraction is $V_{eh}(\bm{x},\bm{y})=-e^2/(4\pi\epsilon)(d^2+|\bm x-\bm y|^2)^{-1/2}$, with $d$ the layer separation and $\varepsilon$ the permittivity of the insulating spacer layer \cite{Ando1982}. Equation \eqref{SI} is expressed in terms of the Hubbard-Stratonovic pair-field $\Delta_\tau(\bm{x},\bm{y})\!=\!V_{eh}(\bm{x},\bm{y})
\langle \Psi_{e,\sigma,\tau}(\bm{x})\Psi_{h,\sigma',\tau}(\bm{y}) \rangle$. 
Since the interlayer Coulomb interaction is nearly spin independent, singlet and triplet excitonic channels are considered  degenerate~\cite{Scammel2023}.  
The interaction is decoupled in the spin-valley singlet interlayer exciton channel ($\sigma\neq\sigma'$)~\cite{Perali2013,Conti2017,Hojlund2023}, allowing us to use a spin-valley scalar order parameter rather than the full spin matrix $\Delta_{\sigma\sigma'}$.
The singlet projection is a simplifying choice and not imposed by any material selection rule.
%

The phase dependence in $S$ is made explicit by performing a spin-independent gauge transformation on the fermionic fields, 
$\Psi_{\lambda,\sigma,\tau}\rightarrow\Psi_{\lambda,\sigma,\tau} e^{i\theta(\bm{x},\tau)/2}$. 
The non-interaction part of the action, Eq.\ \eqref{S0}, transforms as $S_0\rightarrow S_0+S_{\mathrm{ph}}$ where,
\begin{align}
   S_{\mathrm{ph}}&\![\theta] = \sum_{\lambda,\sigma} \int \limits_0^\beta\! \mathrm{d}\tau \int \mathrm{d}\bm{x}    
   ~\Psi^{\dagger}_{\lambda,\sigma,\tau}(\bm{x}) \nonumber \\
   &\!\times\!\!\left( \frac{i}{2}\partial_\tau\theta\!
   +\!\frac{1}{8m^*}(\nabla\theta)^2 
   \!+\!\frac{i}{4m^*}\nabla\theta \!\cdot\! \overset{\leftrightarrow}{\nabla}   
   \right)\! \Psi_{\lambda,\sigma,\tau}(\bm{x})  \label{Sphase} \ ,
\end{align}

with $\nabla\theta\cdot \overset{\leftrightarrow}{\nabla} \Psi_{\lambda,\sigma,\tau}(\bm{x})=\nabla^2\theta\cdot\Psi_{\lambda,\sigma,\tau}(\bm{x})+\nabla\theta \cdot \nabla\Psi_{\lambda,\sigma,\tau}(\bm{x})$.
The interaction part of the action, Eq.\ \eqref{SI}, is left invariant since the pairing field transforms under the gauge transformation as $\Delta_{\tau}(\bm{x},\bm{y})=|\Delta_{\tau}(\bm{x},\bm{y})|e^{i[\theta(\bm{x},\tau)+\theta(\bm{y},\tau)]/2}.$
%
Thus, for a Coulomb interaction, as for a contact interaction, the phase field
$\theta$ enters only through the noninteracting term $S_{\mathrm{ph}}$. 
This occurs for any interaction that is independent of spin and independent of the pair center of mass coordinate, and in the absence of spin-orbit coupling \cite{Devreese2022}. 
The present derivation extends the result for a separable interaction \cite{Benfatto2004} to a nonseparable long-range interaction.

We now separate the action into mean-field and phase-fluctuation contributions, expanding around the superfluid saddle-point \cite{Benfatto2004, Tempere2009, Shi2024, Botelho2006}.
In the hydrodynamic limit  corresponding to the long-wavelength and low-frequency regime, $q\xi \ll 1$ and $\omega \ll 2\Delta$ with $\xi$ the healing length, the Anderson--Bogoliubov branch is acoustic at small momentum, $\omega(q)=c_s q$ \cite{Benfatto2004}.
 The sound velocity $c_s$ is temperature dependent and is given by,
\color{black}
\begin{equation}
c_s(T)=\sqrt{\frac{J_0(T)}{\kappa_0(T)}}\ ,
\label{cseq}
\end{equation}
where $\kappa_0=\kappa/2$ and $\kappa=\left.\frac{\partial n}{\partial\mu}\right|_{T,V}$ is the compressibility \cite{Shi2024, Tempere2009, Benfatto2004}. 
The phase stiffness is,
\begin{equation}
J_0(T) = \frac{n_X}{m_X^*}-\frac{g}{A} \sum_{\mathbf{k}}
\left(\frac{k_x^2}{{m_X^*}^2}\right) n_F(E_{\mathbf{k}},T)\ .
\label{stiffeq}
\end{equation}
$n_F(E_\mathbf{k},T)\!=\!1/(e^{E_\mathbf{k}(T)/T}\!+\!1)$ is the Fermi-Dirac distribution function, and $E_\mathbf{k}(T)\!=\!\sqrt{\epsilon_\mathbf{k}^2+|\Delta_0(\mathbf{k},T)|^2}$ is the superfluid excitation energy with $\epsilon_\mathbf{k}=\mathbf{k}^2/2m^*-\mu$.  $A$ is the area of each layer, $k_x$ is the $x$-component of the vector momentum $\mathbf{k}$. $n_X=n$ and $m_X^*=2m^*$ are the exciton density and effective mass. Since the Coulomb interaction is spin and valley independent, the $\sigma$ index can be omitted, with the spin and valley degeneracy accounted for by a prefactor $g$. 

At zero temperature, Eq.\ \eqref{stiffeq} reduces to $J_0(0)=n/(2m^*)$. This is twice the BCS value for an electron gas with electron density $n$ since in a BCS superconductor two electrons form a single Cooper pair,
whereas in an electron--hole bilayer each electron binds one hole to form one exciton, so $n_X=n$. This one-to-one conversion of carriers into excitons gives the exciton condensate a bare stiffness that is a factor of two larger than for the BCS Cooper-pair condensate. Since $T_{\mathrm{BKT}}$ is proportional to the stiffness, this provides a significant intrinsic advantage for exciton bilayers, allowing substantially higher BKT transition temperatures.
Similarly, the compressibility of the excitons is twice the BCS compressibility.

The stiffness and compressibility depend on the saddle-point values $\Delta_0$ and $\mu$. These are obtained from the gap and number equations 
that follow from minimizing the mean-field contribution $S_{MF}$ of the action at temperature $T$,
\begin{align}
\Delta_0(\mathbf{k},T)&\!=\!-\frac{1}{A}\!\sum_{\mathbf{k}'}\!V_{eh}^{RPA}\!(\mathbf{k}\!-\!\mathbf{k}',T)\frac{\Delta_0(\mathbf{k}',T)\!\tanh\!\left(\!\!\frac{E_{\mathbf{k}'}(T)}{2T}\!\!\right)}{2E_{\mathbf{k}'}(T)} \label{gapeq}\\
    n_{MF}&=\frac{g}{A}\sum_\mathbf{k}\frac{1}{2}\left(1-\frac{\epsilon_\mathbf{k}\tanh\left(\frac{E_{\mathbf{k}}(T)}{2T}\right)}{E_\mathbf{k}(T)}\right) \ .
\end{align}
$V_{eh}^{RPA}$ is the static screening within RPA \cite{Neilson2014} (see Appendix). 
We now include phase fluctuations by decomposing the phase field into
$\theta=\theta_{\mathrm{sw}}+\theta_v$, where $\theta_{\mathrm{sw}}$ denotes
smooth spin-wave fluctuations and $\theta_v$ is the vortex contribution~\cite{Shi2024}.
For the range of densities considered here, $T_{\mathrm{BKT}}$ lies well below the
temperature scale at which spin-waves appreciably affect the density equation so
their contribution can be neglected (for details see the Supplemental material). The transition is therefore controlled by the vortex--antivortex fluctuations.

For $T<T_{\mathrm{BKT}}$, bound vortex--antivortex pairs renormalize the phase stiffness as the temperature increases. At $T=T_{\mathrm{BKT}}$, the pairs unbind, destroying the quasi-long-range order and producing the universal stiffness jump~\cite{Nelson1977}.  As already noted, this is directly reflected in the AB sound velocity $c_s(T)$ (Eq.~\eqref{cseq}). A first estimate of the BKT transition temperature is given by the Nelson--Kosterlitz criterion $T_{\mathrm{BKT}}^{0}=\frac{\pi}{2}\,J_0(T_{\mathrm{BKT}}^{0})$ \cite{Nelson1977}, 

To account for the vortex-induced renormalization of the stiffness, we solve
the Kosterlitz--Thouless RG equations \cite{Nelson1977}: $ \partial_\ell K_\ell^{-1}=4\pi^3 y_\ell^2$ and $\partial_\ell y_\ell=(2-\pi K_\ell)y_\ell$.
$K_\ell(T)=J_\ell(T)/T$  and $y_\ell(T)=\exp(-\mu_{\mathrm{vor},\ell}/T)$ is the vortex
fugacity.  $J_{\ell=0}(T)\equiv J_0(T)$  (Eq.~\eqref{stiffeq}).

The bare vortex-core energy $\mu_{\mathrm{vor},\ell}$ is determined microscopically from the solution of the
Gross--Pitaevskii (GP) equation for a dilute two-dimensional exciton gas. We solve the GP equation using the nonlocal exciton-exciton interaction and obtain the corresponding vortex profile and vortex core energy. 
In the deep BEC regime and for the range of densities considered here (details in the Appendix), we find no appreciable deviation from the standard estimate $\mu_{\mathrm{vor},0}(T)=2.45J_0(T)$, obtained for a 2D Bose gas with purely contact interactions~\cite{Benfatto2007}. We therefore use these values for 
$\mu_{\mathrm{vor},0}(T)$.
The RG-renormalized transition temperature is then,
\begin{equation}
T_{\mathrm{BKT}}^{\mathrm{RG}}
=
\frac{\pi}{2}\,J_{RG}(T_{\mathrm{BKT}}^{\mathrm{RG}})\ ,
\label{stiffRG}
\end{equation}
with $J_{RG}(T)\equiv J_{\ell\to\infty}(T)$ the fully renormalized stiffness.

We now investigate for the first time the properties of the exciton superfluid at finite temperatures. First the gap and number equations are solved in the low-density BEC regime as a function of temperature and then the temperature dependence of the superfluid RPA screening is determined.



%
\begin{figure}[t]
\hspace*{-0.5cm}\includegraphics[trim=0.0cm 0.0cm 0.0cm 0.0cm, clip=true, width=0.47\textwidth]{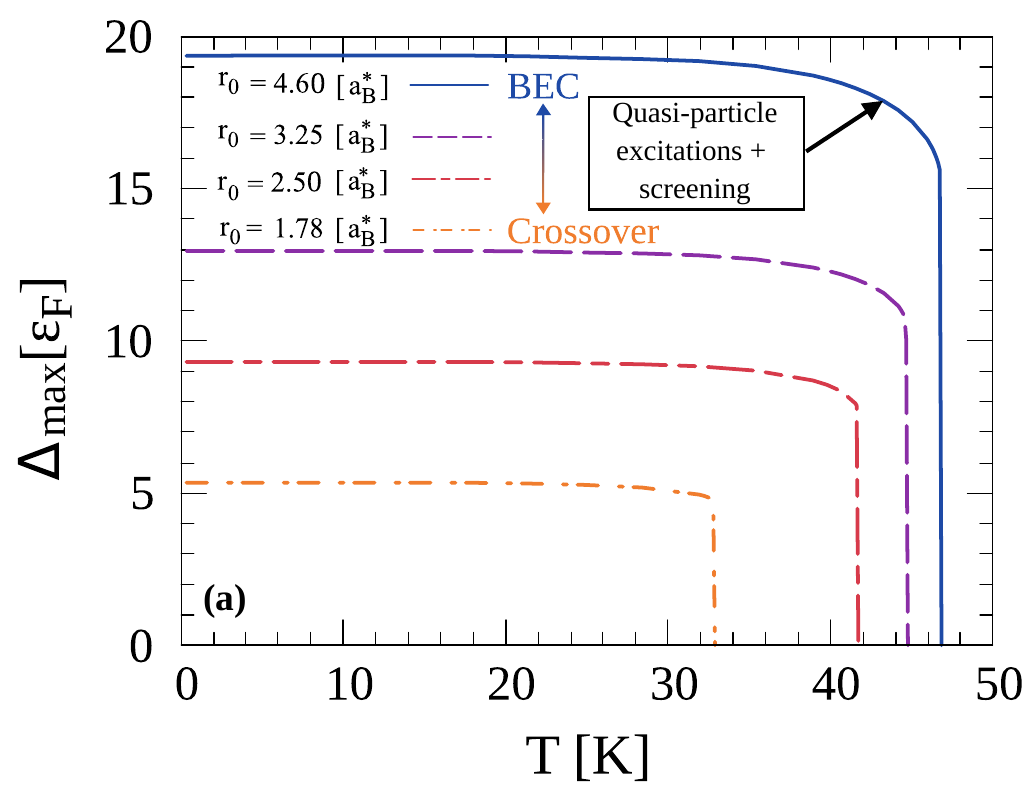}\\
\hspace*{-0.3cm}\includegraphics[trim=0.0cm 0.0cm 0.0cm -1.0cm, clip=true, width=0.47\textwidth]{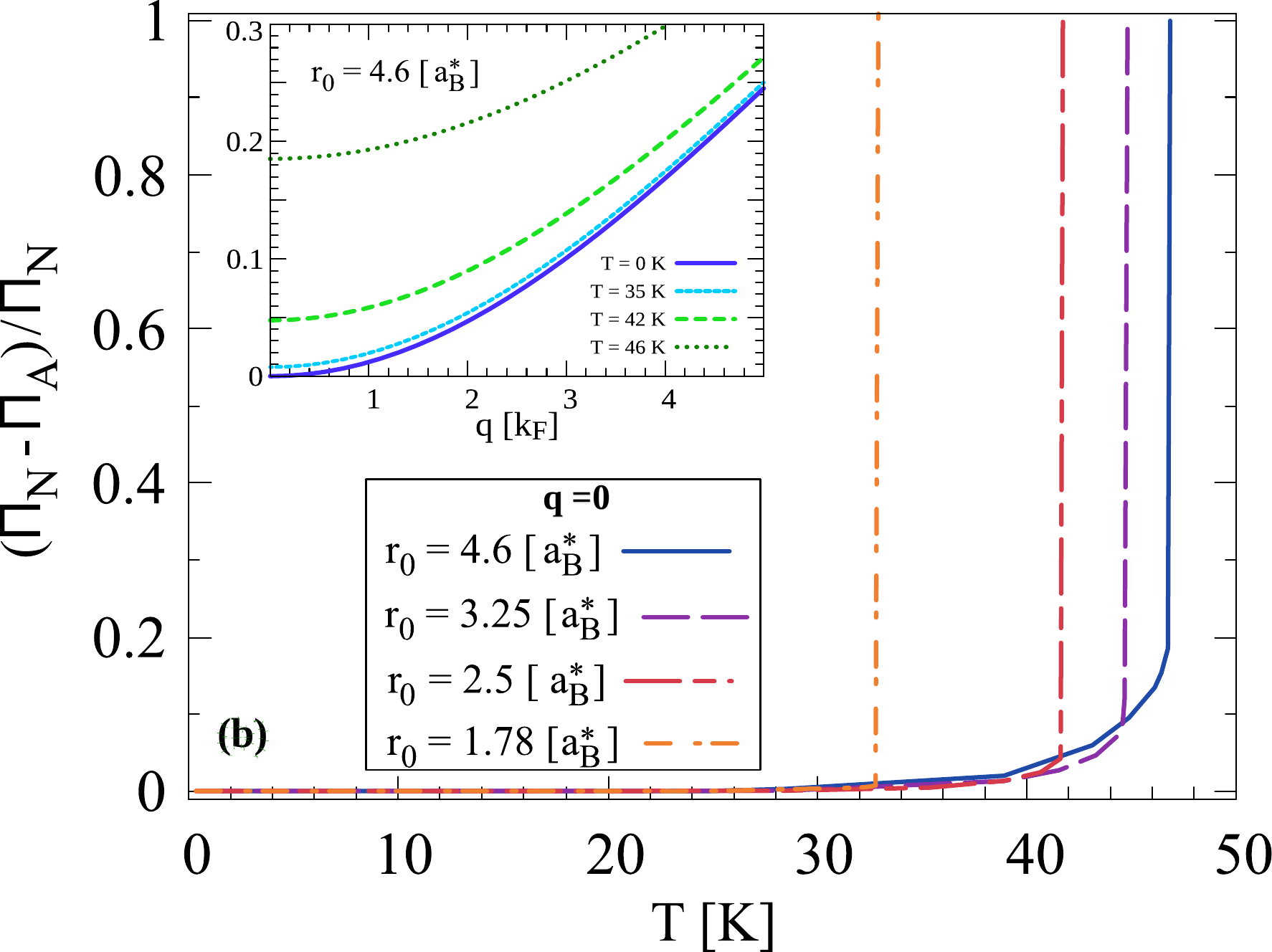}
\caption{(a) Maximum gap energy $\Delta_{max}$ as a function of the temperature $T$. The $r_0$ are the inlayer interparticle distances for different densities. 
(b) Difference between the normal and anomalous polarization functions $\left[\Pi_N-\Pi_A\right]$, normalized to $\Pi_N$,  $\mathbf{q}=0$ as a function of $T$. Inset: $\left[\Pi_N-\Pi_A/\Pi_N\right]$ for $r_0=4.6 a_B^{*}$ as a function of $\mathbf{q}$ for different $T$.}
\label{Fig1}
\end{figure}

Figure~\ref{Fig1}(a) shows the temperature dependence of the maximum gap $\Delta_{\max}$ for double-bilayer graphene (DBG).  We take $m_*=0.045\,m_e$, $g=4$, $\varepsilon=3\varepsilon_0$ \cite{Castro2007,Zou2011}. The effective Bohr radius is $a_B^*=7.9\,\mathrm{nm}$ and the effective Rydberg $Ry^*=30$ meV. The interlayer separation is fixed at  $d=2\,\mathrm{nm}$ and the densities $n$ are labeled by the average inlayer interparticle distances $r_0=1/\sqrt{\pi n}$. We restrict the $r_0>1.78\,a_B^*$, for which the zero-temperature $\Delta_{\max}$ has not yet reached its maximum as a function of density, so Hartree--Fock intralayer contributions can be neglected \cite{Pascucci2024}. 

At low density, $\Delta_{\max}$ decreases smoothly with temperature before collapsing at the mean-field critical temperature $T_c^{\mathrm{MF}}$. Unlike in conventional BCS superconductors, the exciton superfluid mean-field transition is not driven primarily by thermal pair breaking but by the weakening of the electron--hole attraction due to increased screening.
With increasing density, the precursor decrease is progressively lost and the collapse due to superfluid screening becomes sharper. This behavior is controlled by the temperature-dependent superfluid screening. In Fig.~\ref{Fig1}(b), this is shown through the cancellation between the normal and anomalous static polarizations, $\Pi_N(\mathbf{q})-\Pi_A(\mathbf{q})$. The more complete the cancellation for $q\lesssim 2k_F$, the weaker the screening~\cite{Neilson2014}.

At $T=0$, the cancellation is exact at $\mathbf{q}=0$~\cite{Lozovik2012} and, at low density, remains nearly complete up to $q\simeq 2k_F$
(inset of Fig.~\ref{Fig1}(b)), strongly suppressing screening. At finite temperature, thermal pair breaking introduces excited electrons and holes through the pair-breaking terms in Eqs.~\eqref{PiNT}--\eqref{PiAT}, so the cancellation is no longer exact even at $\mathbf{q}=0$. For $r_0=4.6,a_B^*$, the cancellation remains strong up to $T\simeq 35,\mathrm{K}$ and then rapidly weakens as thermally excited carriers enhance screening. This drives the suppression and eventual collapse of $\Delta_{\max}$ at $T_c^{\mathrm{MF}}$.

The density dependence of $T_c^{\mathrm{MF}}$ follows from the ratio $\Delta_{\max}/\varepsilon_F$. In the dilute regime this ratio is large, so thermal quasiparticle excitation is costly and screening turns on only gradually. At higher density, $\Delta_{\max}/\varepsilon_F$ is smaller and more carriers contribute to screening; once the $\mathbf{q}=0$ cancellation is lost, screening is rapidly enhanced over $0<q<2k_F$, producing an abrupt gap collapse.

\begin{figure}[t]
\hspace*{-0.5cm}\includegraphics[width=0.47\textwidth]{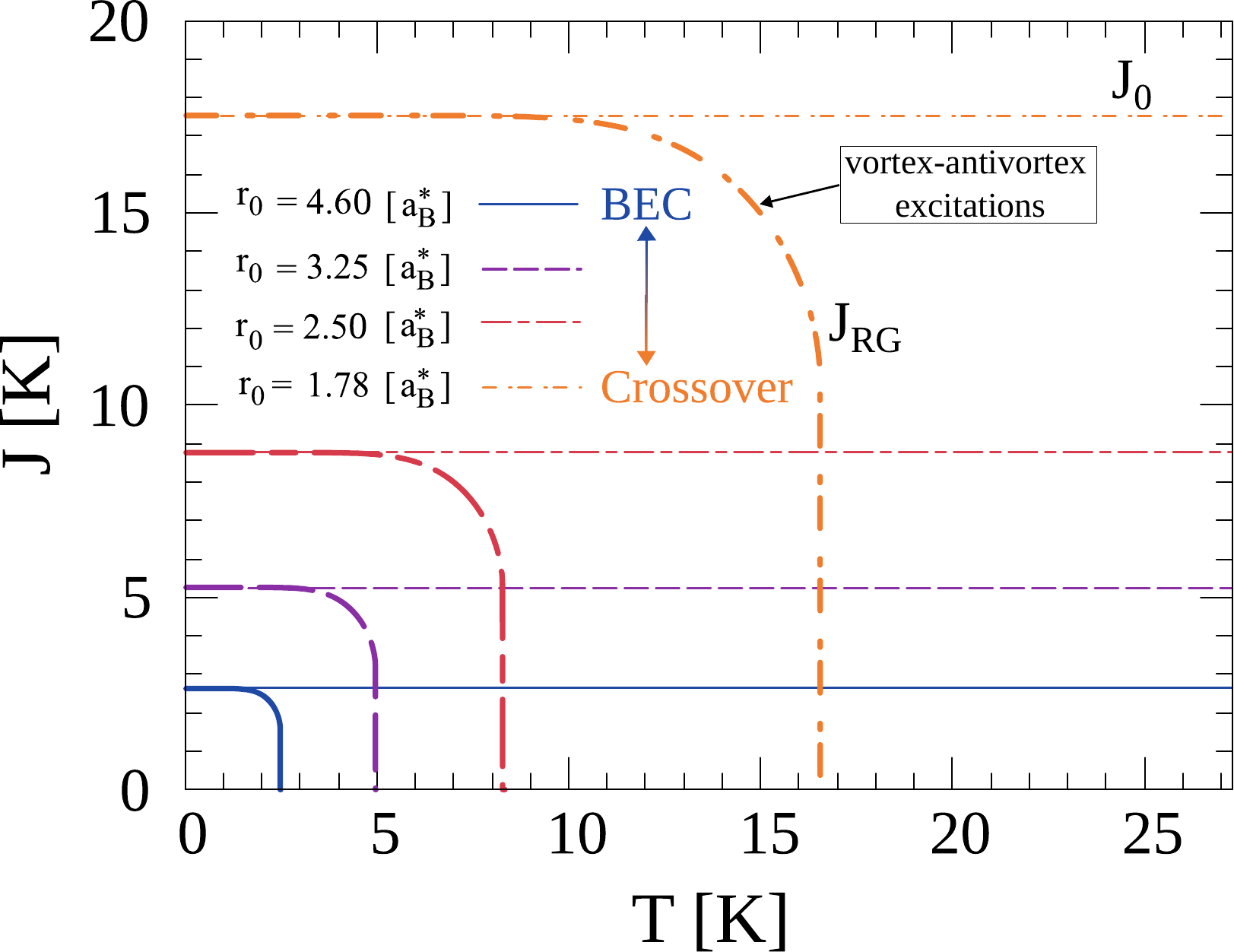}
\caption{Stiffness $J_0(T)$ (horizontal lines) and renormalized stiffness $J_{RG}(T)$ as a function of temperature $T$ for interparticle distances $r_0$, as labeled.
}
\label{stiff}
\end{figure}

In Fig.~\ref{stiff} we compare the bare stiffness $J_0(T)$ (Eq.~\eqref{stiffeq}), with the renormalized stiffness $J_{\mathrm{RG}}(T)$.
At low temperatures, $J_0(T)$ and $J_{\mathrm{RG}}(T)$ coincide since there are a negligible number of thermally excited vortex--antivortex pairs. $J_0(T)$ remains practically temperature independent for all $T<T_c^{\mathrm{MF}}$, reflecting the negligible quasiparticle contribution to Eq.~\eqref{stiffeq}, leaving $J_0\simeq n/(2m^*)$. At  $T=T_c^{\mathrm{MF}}$, $J_0(T)$ then drops suddenly to zero.
In contrast, $J_{\mathrm{RG}}(T)$ is progressively suppressed with increasing temperature due to the proliferation of thermally excited vortex--antivortex pairs in the superfluid.
At the renormalized BKT transition temperature $T_{BKT}^{RG}$, vortex--antivortex unbinding occurs, the stiffness collapses to zero, and the coherent superfluid phase is destroyed.
We can conclude that in this system the vortex--antivortex interaction and the quasi-particle contribution are disentangled in the superfluid.

A central result in  Fig.~\ref{stiff} is that in exciton bilayers, the vortex-renormalization temperature window for $J_{\mathrm{RG}}(T)$ extends over a number of degrees K for both the BEC and BEC-BCS crossover regimes. This is a striking result for a solid-state platform, recalling that in weak-coupled BCS superconductors, the vortex--antivortex fluctuations are usually confined to a mK scale close to the transition~\cite{Weitzel2023, Midei2024}, making the fluctuations very  difficult indeed to isolate experimentally. 

\begin{figure}[t]
 \hspace*{-0.5cm}   \includegraphics[trim=0.0cm 0.0cm 0.0cm 0.0cm, clip=true, width=0.47\textwidth]{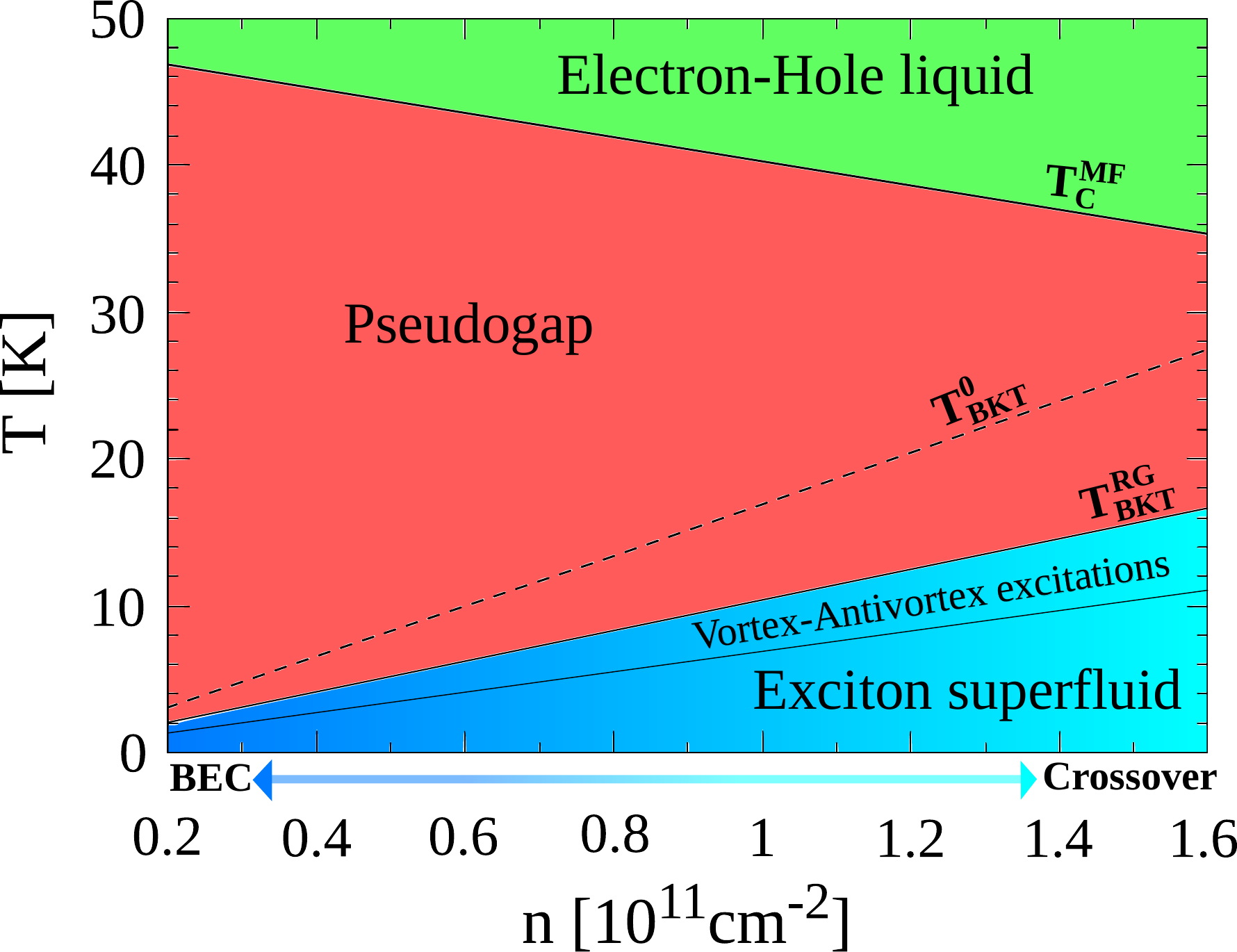}
\caption{Temperature-density ($T,n$) phase-diagram for double-bilayer-graphene. $T_{c}^{MF}$: mean-field critical temperature.   $T_{\mathrm{BKT}}^0$ and  $T_{\mathrm{BKT}}^{\mathrm{RG}}$: BKT transition temperature without and with RG.}
\label{tmftbkt}
\end{figure}
\color{black}
Figure~\ref{tmftbkt} shows the temperature--density phase diagram for DBG. For $T<T_{\mathrm{BKT}}^{\mathrm{RG}}$, 
the system is an exciton superfluid, while for $T>T_c^{\mathrm{MF}}$ it is a normal electron--hole liquid.  
In the intermediate temperature interval,  $T_{\mathrm{BKT}}^{\mathrm{RG}}<T<T_c^{\mathrm{MF}}$, there is still local pairing but unbinding of the vortex--antivortex pairs has destroyed the superfluid phase~\cite{Emery1995,Timusk1999,Lee2009}. We denote this regime in Fig.~\ref{tmftbkt} as the pseudogap.
The temperature-density window over which vortex excitations renormalize the stiffness is also indicated.
$T_{\mathrm{BKT}}^{0}$ is the unrenormalized transition temperature (Eq.\ \eqref{stiffRG}).  
 
The RG flow significantly suppresses $T_{\mathrm{BKT}}^{\mathrm{RG}}$, typically reducing it by a factor of two relative to $T_{\mathrm{BKT}}^{0}$. Thus vortex renormalization is a leading and highly significant correction.
$T_{\mathrm{BKT}}^{\mathrm{RG}}$ increases approximately linearly with density, since over the density range considered 
$T_{\mathrm{BKT}}^{\mathrm{RG}}\ll T_c^{\mathrm{MF}}$. Hence quasiparticle excitations are exponentially suppressed, leaving vortex--antivortex excitations as the dominant mechanism in  renormalizing the stiffness.

\begin{figure}[t]
  \hspace{-0.5cm}  \includegraphics[trim=0.0cm 0.0cm 0.0cm 0.0cm, clip=true, width=0.47\textwidth]{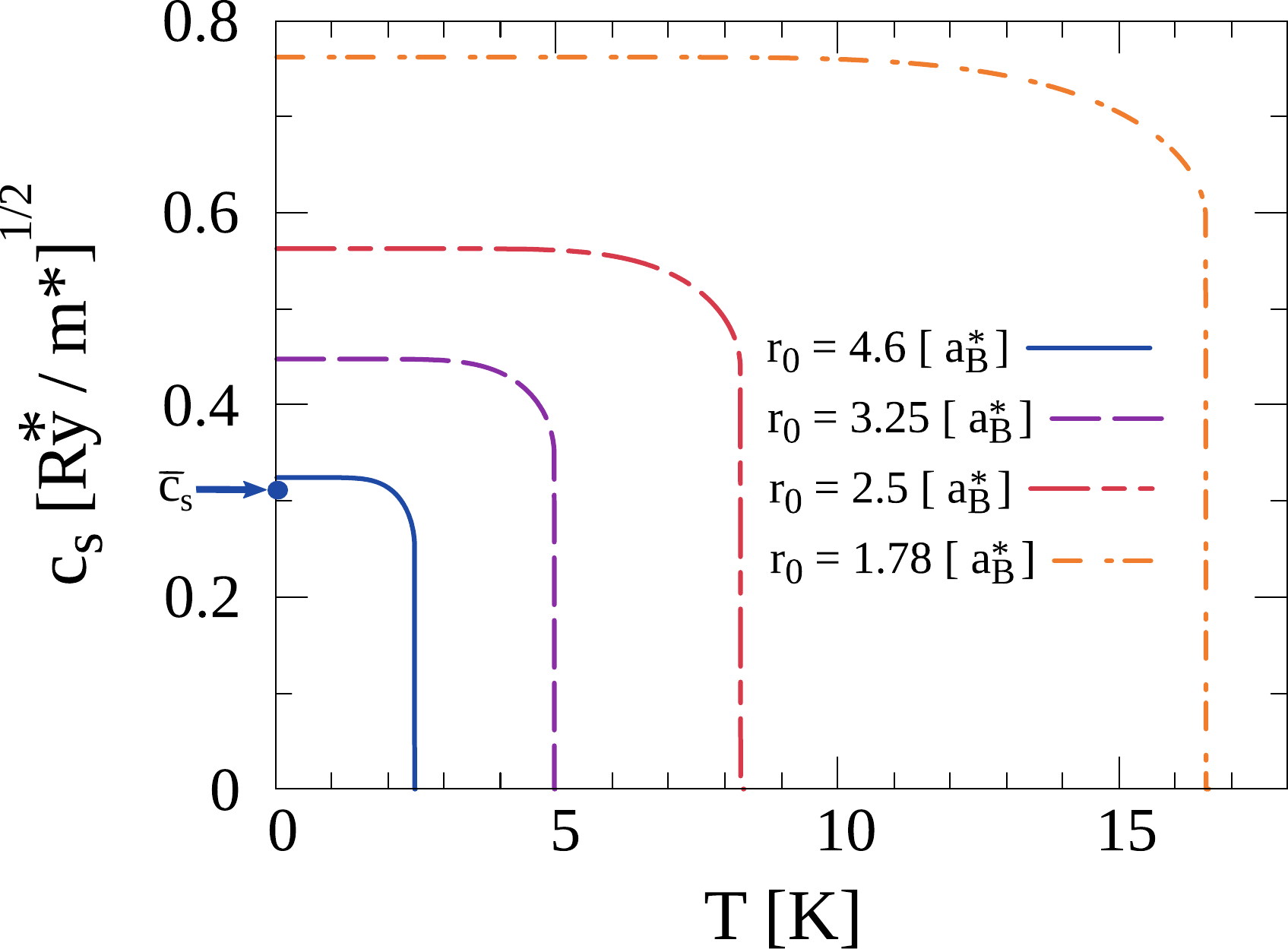}
\caption{Speed of sound $c_s$ for AB mode as a function of temperature $T$ for different interparticle distances $r_0$, as labeled. $\overline{c}_s=\sqrt{2\mu_s+\varepsilon_B/2m^*}$ for $r_0=4.6 a_B^*$, obtained from composite boson theory at $T=0$.}
\label{vf}
\end{figure}

The compressibility $\kappa$ for $T<T_{\mathrm{BKT}}^{\mathrm{RG}}$ is likewise practically temperature
independent, so from Eq.\ \eqref{cseq}, the AB sound velocity $c_s(T)$ rigidly follows $\sqrt{J_{RG}(T)}$, rendering vortex--antivortex fluctuations directly visible from the sound velocity in Fig.~\ref{vf}. 

When the temperature is increased, $c_s$ becomes steadily more suppressed by thermally excited vortex pairs. It drops discontinuously to zero at $T_{\mathrm{BKT}}^{RG}$. With increasing density,  $c_s(T)$ and $T_{\mathrm{BKT}}^{RG}$ both increase as they track $J_{\mathrm{RG}}(T)$ (Fig.~\ref{stiff}).
Also shown in Fig.\ \ref{vf}, is the zero-temperature sound velocity for composite-bosons, $\overline{c}_s=\sqrt{\frac{2\mu_s+\varepsilon_B/2}{m^*}}$ \cite{Salasnich2015} at $r_0=4.6\,a_B^*$,  with $\varepsilon_B$ our calculated exciton binding energy. We note the excellent agreement with our calculated $c_s(T=0)$. 
This confirms the bosonic character of the exciton low-density regime.

The drop ($\sim$ Fermi velocity) of $c_s$ near $T_{\mathrm{BKT}}^{\mathrm{RG}}$ occurs over several Kelvin and makes the vortex-renormalization regime experimentally resolvable.
Since $\kappa$ is almost  temperature independent below $T_{\mathrm{BKT}}^{\mathrm{RG}}$, measuring $c_s^2(T)$ directly tracks $J_{\mathrm{RG}}(T)$, providing access to the renormalized stiffness.
The collapse of $c_s$ is large and is our proposed signature of both the BKT transition and the existence of the exciton superfluid. 

The mode could be probed through density-response techniques, including Bragg spectroscopy~\cite{Combescot2006,Kuhn2020} or through phase-sensitive schemes based on weak coupling to a coherent reservoir~\cite{Kurkjian2019,Valtolina2015}. 

To summarize, the temperature-dependent screening in the superfluid arising from thermal pair-breaking, drives the mean-field transition to the normal electron-hole liquid.
Our temperature--density exciton phase diagram includes a RG vortex--antivortex excitation region which joins the  superfluid, pseudogap, and normal electron--hole liquid phases \cite{Conti2023}. Remarkably, the vortex-antivortex region extends over several degrees Kelvin, in contrast to the few-mK scale typical of solid-state superconductors \cite{Weitzel2023,Midei2024, Furutani2024}. This region, where deviations from mean-field caused by vortex–antivortex pairs in the RG flow, is thus experimentally accessible in exciton bilayers.
The AB sound velocity $c_s(T)$ is steadily suppressed with increasing temperature by the vortex--antivortex excitations. This is followed by a sudden drop to zero at the BKT transition temperature. We demonstrate that the temperature-dependence of the superfluid stiffness, for which magnetic penetration depth measurements are unavailable, is directly obtainable from $c_s(T)$. From this, one can obtain  the temperature window where vortex-antivortex RG excitations are significant. 
\\
\\
{\bf Acknowledgments}
We thank Hadrien Kurkjian and Davide Valentinis for useful discussions. This work was supported by Fonds Wetenschappelijk Onderzoek (FWO) (Grant: 1224225N) and the Provincia Autonoma di Trento.


\begin{thebibliography}{56}%
\makeatletter
\providecommand \@ifxundefined [1]{%
 \@ifx{#1\undefined}
}%
\providecommand \@ifnum [1]{%
 \ifnum #1\expandafter \@firstoftwo
 \else \expandafter \@secondoftwo
 \fi
}%
\providecommand \@ifx [1]{%
 \ifx #1\expandafter \@firstoftwo
 \else \expandafter \@secondoftwo
 \fi
}%
\providecommand \natexlab [1]{#1}%
\providecommand \enquote  [1]{``#1''}%
\providecommand \bibnamefont  [1]{#1}%
\providecommand \bibfnamefont [1]{#1}%
\providecommand \citenamefont [1]{#1}%
\providecommand \href@noop [0]{\@secondoftwo}%
\providecommand \href [0]{\begingroup \@sanitize@url \@href}%
\providecommand \@href[1]{\@@startlink{#1}\@@href}%
\providecommand \@@href[1]{\endgroup#1\@@endlink}%
\providecommand \@sanitize@url [0]{\catcode `\\12\catcode `\$12\catcode
  `\&12\catcode `\#12\catcode `\^12\catcode `\_12\catcode `\%12\relax}%
\providecommand \@@startlink[1]{}%
\providecommand \@@endlink[0]{}%
\providecommand \url  [0]{\begingroup\@sanitize@url \@url }%
\providecommand \@url [1]{\endgroup\@href {#1}{\urlprefix }}%
\providecommand \urlprefix  [0]{URL }%
\providecommand \Eprint [0]{\href }%
\providecommand \doibase [0]{https://doi.org/}%
\providecommand \selectlanguage [0]{\@gobble}%
\providecommand \bibinfo  [0]{\@secondoftwo}%
\providecommand \bibfield  [0]{\@secondoftwo}%
\providecommand \translation [1]{[#1]}%
\providecommand \BibitemOpen [0]{}%
\providecommand \bibitemStop [0]{}%
\providecommand \bibitemNoStop [0]{.\EOS\space}%
\providecommand \EOS [0]{\spacefactor3000\relax}%
\providecommand \BibitemShut  [1]{\csname bibitem#1\endcsname}%
\let\auto@bib@innerbib\@empty
\bibitem [{\citenamefont {Burg}\ \emph {et~al.}(2018)\citenamefont {Burg},
  \citenamefont {Prasad}, \citenamefont {Kim}, \citenamefont {Taniguchi},
  \citenamefont {Watanabe}, \citenamefont {MacDonald}, \citenamefont
  {Register},\ and\ \citenamefont {Tutuc}}]{Burg2018}%
  \BibitemOpen
  \bibfield  {author} {\bibinfo {author} {\bibfnamefont {G.~W.}\ \bibnamefont
  {Burg}}, \bibinfo {author} {\bibfnamefont {N.}~\bibnamefont {Prasad}},
  \bibinfo {author} {\bibfnamefont {K.}~\bibnamefont {Kim}}, \bibinfo {author}
  {\bibfnamefont {T.}~\bibnamefont {Taniguchi}}, \bibinfo {author}
  {\bibfnamefont {K.}~\bibnamefont {Watanabe}}, \bibinfo {author}
  {\bibfnamefont {A.~H.}\ \bibnamefont {MacDonald}}, \bibinfo {author}
  {\bibfnamefont {L.~F.}\ \bibnamefont {Register}},\ and\ \bibinfo {author}
  {\bibfnamefont {E.}~\bibnamefont {Tutuc}},\ }\bibfield  {title} {\bibinfo
  {title} {Strongly enhanced tunneling at total charge neutrality in
  double-bilayer graphene-{${\mathrm{WSe}}_{2}$} heterostructures},\ }\href
  {https://doi.org/10.1103/PhysRevLett.120.177702} {\bibfield  {journal}
  {\bibinfo  {journal} {Phys. Rev. Lett.}\ }\textbf {\bibinfo {volume} {120}},\
  \bibinfo {pages} {177702} (\bibinfo {year} {2018})}\BibitemShut {NoStop}%
\bibitem [{\citenamefont {Wang}\ \emph {et~al.}(2019)\citenamefont {Wang},
  \citenamefont {Rhodes}, \citenamefont {Watanabe}, \citenamefont {Taniguchi},
  \citenamefont {Hone}, \citenamefont {Shan},\ and\ \citenamefont
  {Mak}}]{Wang2019}%
  \BibitemOpen
  \bibfield  {author} {\bibinfo {author} {\bibfnamefont {Z.}~\bibnamefont
  {Wang}}, \bibinfo {author} {\bibfnamefont {D.~A.}\ \bibnamefont {Rhodes}},
  \bibinfo {author} {\bibfnamefont {K.}~\bibnamefont {Watanabe}}, \bibinfo
  {author} {\bibfnamefont {T.}~\bibnamefont {Taniguchi}}, \bibinfo {author}
  {\bibfnamefont {J.~C.}\ \bibnamefont {Hone}}, \bibinfo {author}
  {\bibfnamefont {J.}~\bibnamefont {Shan}},\ and\ \bibinfo {author}
  {\bibfnamefont {K.~F.}\ \bibnamefont {Mak}},\ }\bibfield  {title} {\bibinfo
  {title} {Evidence of high-temperature exciton condensation in two-dimensional
  atomic double layers},\ }\href {https://doi.org/10.1038/s41586-019-1591-7}
  {\bibfield  {journal} {\bibinfo  {journal} {Nature (London)}\ }\textbf
  {\bibinfo {volume} {574}},\ \bibinfo {pages} {76} (\bibinfo {year}
  {2019})}\BibitemShut {NoStop}%
\bibitem [{\citenamefont {Gu}\ \emph {et~al.}(2022)\citenamefont {Gu},
  \citenamefont {Ma}, \citenamefont {Liu}, \citenamefont {Watanabe},
  \citenamefont {Taniguchi}, \citenamefont {Hone}, \citenamefont {Shan},\ and\
  \citenamefont {Mak}}]{Gu2022}%
  \BibitemOpen
  \bibfield  {author} {\bibinfo {author} {\bibfnamefont {J.}~\bibnamefont
  {Gu}}, \bibinfo {author} {\bibfnamefont {L.}~\bibnamefont {Ma}}, \bibinfo
  {author} {\bibfnamefont {S.}~\bibnamefont {Liu}}, \bibinfo {author}
  {\bibfnamefont {K.}~\bibnamefont {Watanabe}}, \bibinfo {author}
  {\bibfnamefont {T.}~\bibnamefont {Taniguchi}}, \bibinfo {author}
  {\bibfnamefont {J.~C.}\ \bibnamefont {Hone}}, \bibinfo {author}
  {\bibfnamefont {J.}~\bibnamefont {Shan}},\ and\ \bibinfo {author}
  {\bibfnamefont {K.~F.}\ \bibnamefont {Mak}},\ }\bibfield  {title} {\bibinfo
  {title} {Dipolar excitonic insulator in a moir{\'e} lattice},\ }\href
  {https://doi.org/10.1038/s41567-022-01532-z} {\bibfield  {journal} {\bibinfo
  {journal} {Nat. Phys.}\ }\textbf {\bibinfo {volume} {18}},\ \bibinfo {pages}
  {395} (\bibinfo {year} {2022})}\BibitemShut {NoStop}%
\bibitem [{\citenamefont {Ma}\ \emph {et~al.}(2021)\citenamefont {Ma},
  \citenamefont {Nguyen}, \citenamefont {Wang}, \citenamefont {Zeng},
  \citenamefont {Watanabe}, \citenamefont {Taniguchi}, \citenamefont
  {MacDonald}, \citenamefont {Mak},\ and\ \citenamefont {Shan}}]{Ma2021}%
  \BibitemOpen
  \bibfield  {author} {\bibinfo {author} {\bibfnamefont {L.}~\bibnamefont
  {Ma}}, \bibinfo {author} {\bibfnamefont {P.~X.}\ \bibnamefont {Nguyen}},
  \bibinfo {author} {\bibfnamefont {Z.}~\bibnamefont {Wang}}, \bibinfo {author}
  {\bibfnamefont {Y.}~\bibnamefont {Zeng}}, \bibinfo {author} {\bibfnamefont
  {K.}~\bibnamefont {Watanabe}}, \bibinfo {author} {\bibfnamefont
  {T.}~\bibnamefont {Taniguchi}}, \bibinfo {author} {\bibfnamefont {A.~H.}\
  \bibnamefont {MacDonald}}, \bibinfo {author} {\bibfnamefont {K.~F.}\
  \bibnamefont {Mak}},\ and\ \bibinfo {author} {\bibfnamefont {J.}~\bibnamefont
  {Shan}},\ }\bibfield  {title} {\bibinfo {title} {Strongly correlated
  excitonic insulator in atomic double layers},\ }\href
  {https://doi.org/10.1038/s41586-021-03947-9} {\bibfield  {journal} {\bibinfo
  {journal} {Nature (London)}\ }\textbf {\bibinfo {volume} {598}},\ \bibinfo
  {pages} {585} (\bibinfo {year} {2021})}\BibitemShut {NoStop}%
\bibitem [{\citenamefont {Nguyen}\ \emph {et~al.}(2025)\citenamefont {Nguyen},
  \citenamefont {Ma}, \citenamefont {Chaturvedi}, \citenamefont {Watanabe},
  \citenamefont {Taniguchi}, \citenamefont {Shan},\ and\ \citenamefont
  {Mak}}]{Nguyen2025}%
  \BibitemOpen
  \bibfield  {author} {\bibinfo {author} {\bibfnamefont {P.~X.}\ \bibnamefont
  {Nguyen}}, \bibinfo {author} {\bibfnamefont {L.}~\bibnamefont {Ma}}, \bibinfo
  {author} {\bibfnamefont {R.}~\bibnamefont {Chaturvedi}}, \bibinfo {author}
  {\bibfnamefont {K.}~\bibnamefont {Watanabe}}, \bibinfo {author}
  {\bibfnamefont {T.}~\bibnamefont {Taniguchi}}, \bibinfo {author}
  {\bibfnamefont {J.}~\bibnamefont {Shan}},\ and\ \bibinfo {author}
  {\bibfnamefont {K.~F.}\ \bibnamefont {Mak}},\ }\bibfield  {title} {\bibinfo
  {title} {Perfect coulomb drag in a dipolar excitonic insulator},\ }\href
  {https://doi.org/10.1126/science.adl1829} {\bibfield  {journal} {\bibinfo
  {journal} {Science}\ }\textbf {\bibinfo {volume} {388}},\ \bibinfo {pages}
  {274} (\bibinfo {year} {2025})}\BibitemShut {NoStop}%
\bibitem [{\citenamefont {Perali}\ \emph {et~al.}(2013)\citenamefont {Perali},
  \citenamefont {Neilson},\ and\ \citenamefont {Hamilton}}]{Perali2013}%
  \BibitemOpen
  \bibfield  {author} {\bibinfo {author} {\bibfnamefont {A.}~\bibnamefont
  {Perali}}, \bibinfo {author} {\bibfnamefont {D.}~\bibnamefont {Neilson}},\
  and\ \bibinfo {author} {\bibfnamefont {A.~R.}\ \bibnamefont {Hamilton}},\
  }\bibfield  {title} {\bibinfo {title} {High-temperature superfluidity in
  double-bilayer graphene},\ }\href
  {https://doi.org/10.1103/PhysRevLett.110.146803} {\bibfield  {journal}
  {\bibinfo  {journal} {Phys. Rev. Lett.}\ }\textbf {\bibinfo {volume} {110}},\
  \bibinfo {pages} {146803} (\bibinfo {year} {2013})}\BibitemShut {NoStop}%
\bibitem [{\citenamefont {Tutuc}\ \emph {et~al.}(2004)\citenamefont {Tutuc},
  \citenamefont {Shayegan},\ and\ \citenamefont {Huse}}]{Tutuc2004}%
  \BibitemOpen
  \bibfield  {author} {\bibinfo {author} {\bibfnamefont {E.}~\bibnamefont
  {Tutuc}}, \bibinfo {author} {\bibfnamefont {M.}~\bibnamefont {Shayegan}},\
  and\ \bibinfo {author} {\bibfnamefont {D.~A.}\ \bibnamefont {Huse}},\
  }\bibfield  {title} {\bibinfo {title} {Counterflow measurements in strongly
  correlated {G}a{A}s hole bilayers: {E}vidence for electron-hole pairing},\
  }\href {https://doi.org/10.1103/PhysRevLett.93.036802} {\bibfield  {journal}
  {\bibinfo  {journal} {Phys. Rev. Lett.}\ }\textbf {\bibinfo {volume} {93}},\
  \bibinfo {pages} {036802} (\bibinfo {year} {2004})}\BibitemShut {NoStop}%
\bibitem [{\citenamefont {Su}\ and\ \citenamefont {MacDonald}(2008)}]{Su2008}%
  \BibitemOpen
  \bibfield  {author} {\bibinfo {author} {\bibfnamefont {J.-J.}\ \bibnamefont
  {Su}}\ and\ \bibinfo {author} {\bibfnamefont {A.~H.}\ \bibnamefont
  {MacDonald}},\ }\bibfield  {title} {\bibinfo {title} {How to make a bilayer
  exciton condensate flow},\ }\href {https://doi.org/10.1038/nphys1055}
  {\bibfield  {journal} {\bibinfo  {journal} {Nat. Phys.}\ }\textbf {\bibinfo
  {volume} {4}},\ \bibinfo {pages} {799} (\bibinfo {year} {2008})}\BibitemShut
  {NoStop}%
\bibitem [{\citenamefont {Nandi}\ \emph {et~al.}(2012)\citenamefont {Nandi},
  \citenamefont {Finck}, \citenamefont {Eisenstein}, \citenamefont {Pfeiffer},\
  and\ \citenamefont {West}}]{Nandi2012}%
  \BibitemOpen
  \bibfield  {author} {\bibinfo {author} {\bibfnamefont {D.}~\bibnamefont
  {Nandi}}, \bibinfo {author} {\bibfnamefont {A.~D.~K.}\ \bibnamefont {Finck}},
  \bibinfo {author} {\bibfnamefont {J.~P.}\ \bibnamefont {Eisenstein}},
  \bibinfo {author} {\bibfnamefont {L.~N.}\ \bibnamefont {Pfeiffer}},\ and\
  \bibinfo {author} {\bibfnamefont {K.~W.}\ \bibnamefont {West}},\ }\bibfield
  {title} {\bibinfo {title} {Exciton condensation and perfect {C}oulomb drag},\
  }\href {https://doi.org/10.1038/nature11302} {\bibfield  {journal} {\bibinfo
  {journal} {Nature}\ }\textbf {\bibinfo {volume} {488}},\ \bibinfo {pages}
  {481} (\bibinfo {year} {2012})}\BibitemShut {NoStop}%
\bibitem [{\citenamefont {Pieri}\ \emph {et~al.}(2007)\citenamefont {Pieri},
  \citenamefont {Neilson},\ and\ \citenamefont {Strinati}}]{Pieri2007}%
  \BibitemOpen
  \bibfield  {author} {\bibinfo {author} {\bibfnamefont {P.}~\bibnamefont
  {Pieri}}, \bibinfo {author} {\bibfnamefont {D.}~\bibnamefont {Neilson}},\
  and\ \bibinfo {author} {\bibfnamefont {G.~C.}\ \bibnamefont {Strinati}},\
  }\bibfield  {title} {\bibinfo {title} {Effects of density imbalance on the
  {BCS-BEC} crossover in semiconductor electron-hole bilayers},\ }\href
  {https://doi.org/10.1103/PhysRevB.75.113301} {\bibfield  {journal} {\bibinfo
  {journal} {Phys. Rev. B}\ }\textbf {\bibinfo {volume} {75}},\ \bibinfo
  {pages} {113301} (\bibinfo {year} {2007})}\BibitemShut {NoStop}%
\bibitem [{\citenamefont {Salasnich}\ \emph {et~al.}(2013)\citenamefont
  {Salasnich}, \citenamefont {Marchetti},\ and\ \citenamefont
  {Toigo}}]{Salasnich2013}%
  \BibitemOpen
  \bibfield  {author} {\bibinfo {author} {\bibfnamefont {L.}~\bibnamefont
  {Salasnich}}, \bibinfo {author} {\bibfnamefont {P.~A.}\ \bibnamefont
  {Marchetti}},\ and\ \bibinfo {author} {\bibfnamefont {F.}~\bibnamefont
  {Toigo}},\ }\bibfield  {title} {\bibinfo {title} {Superfluidity, sound
  velocity, and quasicondensation in the two-dimensional {BCS-BEC} crossover},\
  }\href {https://doi.org/10.1103/PhysRevA.88.053612} {\bibfield  {journal}
  {\bibinfo  {journal} {Phys. Rev. A}\ }\textbf {\bibinfo {volume} {88}},\
  \bibinfo {pages} {053612} (\bibinfo {year} {2013})}\BibitemShut {NoStop}%
\bibitem [{\citenamefont {L\'opez~R\'{\i}os}\ \emph {et~al.}(2018)\citenamefont
  {L\'opez~R\'{\i}os}, \citenamefont {Perali}, \citenamefont {Needs},\ and\
  \citenamefont {Neilson}}]{LopezRios2018}%
  \BibitemOpen
  \bibfield  {author} {\bibinfo {author} {\bibfnamefont {P.}~\bibnamefont
  {L\'opez~R\'{\i}os}}, \bibinfo {author} {\bibfnamefont {A.}~\bibnamefont
  {Perali}}, \bibinfo {author} {\bibfnamefont {R.~J.}\ \bibnamefont {Needs}},\
  and\ \bibinfo {author} {\bibfnamefont {D.}~\bibnamefont {Neilson}},\
  }\bibfield  {title} {\bibinfo {title} {Evidence from quantum {M}onte {C}arlo
  simulations of large-gap superfluidity and {BCS}-{BEC} crossover in double
  electron-hole layers},\ }\href
  {https://doi.org/10.1103/PhysRevLett.120.177701} {\bibfield  {journal}
  {\bibinfo  {journal} {Phys. Rev. Lett.}\ }\textbf {\bibinfo {volume} {120}},\
  \bibinfo {pages} {177701} (\bibinfo {year} {2018})}\BibitemShut {NoStop}%
\bibitem [{\citenamefont {Spielman}\ \emph {et~al.}(2000)\citenamefont
  {Spielman}, \citenamefont {Eisenstein}, \citenamefont {Pfeiffer},\ and\
  \citenamefont {West}}]{Spielman2000}%
  \BibitemOpen
  \bibfield  {author} {\bibinfo {author} {\bibfnamefont {I.~B.}\ \bibnamefont
  {Spielman}}, \bibinfo {author} {\bibfnamefont {J.~P.}\ \bibnamefont
  {Eisenstein}}, \bibinfo {author} {\bibfnamefont {L.~N.}\ \bibnamefont
  {Pfeiffer}},\ and\ \bibinfo {author} {\bibfnamefont {K.~W.}\ \bibnamefont
  {West}},\ }\bibfield  {title} {\bibinfo {title} {Resonantly enhanced
  tunneling in a double layer quantum {H}all ferromagnet},\ }\href
  {https://doi.org/10.1103/PhysRevLett.84.5808} {\bibfield  {journal} {\bibinfo
   {journal} {Phys. Rev. Lett.}\ }\textbf {\bibinfo {volume} {84}},\ \bibinfo
  {pages} {5808} (\bibinfo {year} {2000})}\BibitemShut {NoStop}%
\bibitem [{\citenamefont {Narozhny}\ and\ \citenamefont
  {Levchenko}(2016)}]{Narozhny2016}%
  \BibitemOpen
  \bibfield  {author} {\bibinfo {author} {\bibfnamefont {B.~N.}\ \bibnamefont
  {Narozhny}}\ and\ \bibinfo {author} {\bibfnamefont {A.}~\bibnamefont
  {Levchenko}},\ }\bibfield  {title} {\bibinfo {title} {Coulomb drag},\ }\href
  {https://doi.org/10.1103/RevModPhys.88.025003} {\bibfield  {journal}
  {\bibinfo  {journal} {Rev. Mod. Phys.}\ }\textbf {\bibinfo {volume} {88}},\
  \bibinfo {pages} {025003} (\bibinfo {year} {2016})}\BibitemShut {NoStop}%
\bibitem [{\citenamefont {Li}\ \emph {et~al.}(2017)\citenamefont {Li},
  \citenamefont {Taniguchi}, \citenamefont {Watanabe}, \citenamefont {Hone},\
  and\ \citenamefont {Dean}}]{Li2017}%
  \BibitemOpen
  \bibfield  {author} {\bibinfo {author} {\bibfnamefont {J.~I.~A.}\
  \bibnamefont {Li}}, \bibinfo {author} {\bibfnamefont {T.}~\bibnamefont
  {Taniguchi}}, \bibinfo {author} {\bibfnamefont {K.}~\bibnamefont {Watanabe}},
  \bibinfo {author} {\bibfnamefont {J.}~\bibnamefont {Hone}},\ and\ \bibinfo
  {author} {\bibfnamefont {C.}~\bibnamefont {Dean}},\ }\bibfield  {title}
  {\bibinfo {title} {Excitonic superfluid phase in double bilayer graphene},\
  }\href {https://www.nature.com/articles/nphys4140} {\bibfield  {journal}
  {\bibinfo  {journal} {Nat. Phys.}\ }\textbf {\bibinfo {volume} {13}},\
  \bibinfo {pages} {751} (\bibinfo {year} {2017})}\BibitemShut {NoStop}%
\bibitem [{\citenamefont {Liu}\ \emph {et~al.}(2022)\citenamefont {Liu},
  \citenamefont {Li}, \citenamefont {Watanabe}, \citenamefont {Taniguchi},
  \citenamefont {Hone}, \citenamefont {Halperin}, \citenamefont {Kim},\ and\
  \citenamefont {Dean}}]{Liu2022}%
  \BibitemOpen
  \bibfield  {author} {\bibinfo {author} {\bibfnamefont {X.}~\bibnamefont
  {Liu}}, \bibinfo {author} {\bibfnamefont {J.~I.~A.}\ \bibnamefont {Li}},
  \bibinfo {author} {\bibfnamefont {K.}~\bibnamefont {Watanabe}}, \bibinfo
  {author} {\bibfnamefont {T.}~\bibnamefont {Taniguchi}}, \bibinfo {author}
  {\bibfnamefont {J.}~\bibnamefont {Hone}}, \bibinfo {author} {\bibfnamefont
  {B.~I.}\ \bibnamefont {Halperin}}, \bibinfo {author} {\bibfnamefont
  {P.}~\bibnamefont {Kim}},\ and\ \bibinfo {author} {\bibfnamefont {C.~R.}\
  \bibnamefont {Dean}},\ }\bibfield  {title} {\bibinfo {title} {Crossover
  between strongly coupled and weakly coupled exciton superfluids},\ }\href
  {https://doi.org/10.1126/science.abg1110} {\bibfield  {journal} {\bibinfo
  {journal} {Science}\ }\textbf {\bibinfo {volume} {375}},\ \bibinfo {pages}
  {205} (\bibinfo {year} {2022})}\BibitemShut {NoStop}%
\bibitem [{\citenamefont {Qi}\ \emph {et~al.}(2026)\citenamefont {Qi},
  \citenamefont {Li}, \citenamefont {Nie}, \citenamefont {Xia}, \citenamefont
  {Kim}, \citenamefont {Lim}, \citenamefont {Xie}, \citenamefont {Taniguchi},
  \citenamefont {Watanabe}, \citenamefont {Crommie}, \citenamefont
  {MacDonald},\ and\ \citenamefont {Wang}}]{Qi2026}%
  \BibitemOpen
  \bibfield  {author} {\bibinfo {author} {\bibfnamefont {R.}~\bibnamefont
  {Qi}}, \bibinfo {author} {\bibfnamefont {Q.}~\bibnamefont {Li}}, \bibinfo
  {author} {\bibfnamefont {J.}~\bibnamefont {Nie}}, \bibinfo {author}
  {\bibfnamefont {R.}~\bibnamefont {Xia}}, \bibinfo {author} {\bibfnamefont
  {H.}~\bibnamefont {Kim}}, \bibinfo {author} {\bibfnamefont {H.}~\bibnamefont
  {Lim}}, \bibinfo {author} {\bibfnamefont {J.}~\bibnamefont {Xie}}, \bibinfo
  {author} {\bibfnamefont {T.}~\bibnamefont {Taniguchi}}, \bibinfo {author}
  {\bibfnamefont {K.}~\bibnamefont {Watanabe}}, \bibinfo {author}
  {\bibfnamefont {M.~F.}\ \bibnamefont {Crommie}}, \bibinfo {author}
  {\bibfnamefont {A.~H.}\ \bibnamefont {MacDonald}},\ and\ \bibinfo {author}
  {\bibfnamefont {F.}~\bibnamefont {Wang}},\ }\bibfield  {title} {\bibinfo
  {title} {Two-component exciton condensates in an electron--hole bilayer},\
  }\bibfield  {journal} {\bibinfo  {journal} {Nature}\ }\href
  {https://doi.org/10.1038/s41586-026-10636-y} {10.1038/s41586-026-10636-y}
  (\bibinfo {year} {2026})\BibitemShut {NoStop}%
\bibitem [{\citenamefont {Berezinsky}(1972)}]{Berezinsky1972}%
  \BibitemOpen
  \bibfield  {author} {\bibinfo {author} {\bibfnamefont {V.~L.}\ \bibnamefont
  {Berezinsky}},\ }\bibfield  {title} {\bibinfo {title} {Destruction of
  long-range order in one-dimensional and two-dimensional systems possessing a
  continuous symmetry group. {II.} quantum systems.},\ }\href
  {http://www.jetp.ac.ru/cgi-bin/e/index/e/34/3/p610?a=list} {\bibfield
  {journal} {\bibinfo  {journal} {Sov. Phys. JETP}\ }\textbf {\bibinfo {volume}
  {34}},\ \bibinfo {pages} {610} (\bibinfo {year} {1972})},\ \bibinfo {note}
  {(Zh. Eksp. Teor. Fiz. \textbf{61}, 1144 (1972))}\BibitemShut {NoStop}%
\bibitem [{\citenamefont {Kosterlitz}\ and\ \citenamefont
  {Thouless}(1973)}]{Kosterlitz1973}%
  \BibitemOpen
  \bibfield  {author} {\bibinfo {author} {\bibfnamefont {J.~M.}\ \bibnamefont
  {Kosterlitz}}\ and\ \bibinfo {author} {\bibfnamefont {D.~J.}\ \bibnamefont
  {Thouless}},\ }\bibfield  {title} {\bibinfo {title} {Ordering, metastability
  and phase transitions in two-dimensional systems},\ }\href
  {https://doi.org/10.1088/0022-3719/6/7/010} {\bibfield  {journal} {\bibinfo
  {journal} {J. Phys. C: Solid State}\ }\textbf {\bibinfo {volume} {6}},\
  \bibinfo {pages} {1181} (\bibinfo {year} {1973})}\BibitemShut {NoStop}%
\bibitem [{\citenamefont {Van~Loon}\ and\ \citenamefont {S\'a~de
  Melo}(2023)}]{VanLoon2023}%
  \BibitemOpen
  \bibfield  {author} {\bibinfo {author} {\bibfnamefont {S.}~\bibnamefont
  {Van~Loon}}\ and\ \bibinfo {author} {\bibfnamefont {C.~A.~R.}\ \bibnamefont
  {S\'a~de Melo}},\ }\bibfield  {title} {\bibinfo {title} {Effects of quantum
  fluctuations on the low-energy collective modes of two-dimensional superfluid
  {F}ermi gases from the {BCS} to the {B}ose limit},\ }\href
  {https://doi.org/10.1103/PhysRevLett.131.113001} {\bibfield  {journal}
  {\bibinfo  {journal} {Phys. Rev. Lett.}\ }\textbf {\bibinfo {volume} {131}},\
  \bibinfo {pages} {113001} (\bibinfo {year} {2023})}\BibitemShut {NoStop}%
\bibitem [{\citenamefont {Nelson}\ and\ \citenamefont
  {Kosterlitz}(1977)}]{Nelson1977}%
  \BibitemOpen
  \bibfield  {author} {\bibinfo {author} {\bibfnamefont {D.~R.}\ \bibnamefont
  {Nelson}}\ and\ \bibinfo {author} {\bibfnamefont {J.~M.}\ \bibnamefont
  {Kosterlitz}},\ }\bibfield  {title} {\bibinfo {title} {Universal jump in the
  superfluid density of two-dimensional superfluids},\ }\href
  {https://doi.org/10.1103/PhysRevLett.39.1201} {\bibfield  {journal} {\bibinfo
   {journal} {Phys. Rev. Lett.}\ }\textbf {\bibinfo {volume} {39}},\ \bibinfo
  {pages} {1201} (\bibinfo {year} {1977})}\BibitemShut {NoStop}%
\bibitem [{\citenamefont {Anderson}(1958)}]{Anderson1958}%
  \BibitemOpen
  \bibfield  {author} {\bibinfo {author} {\bibfnamefont {P.~W.}\ \bibnamefont
  {Anderson}},\ }\bibfield  {title} {\bibinfo {title} {Random--phase
  approximation in the theory of superconductivity},\ }\href
  {https://doi.org/10.1103/PhysRev.112.1900} {\bibfield  {journal} {\bibinfo
  {journal} {Phys. Rev.}\ }\textbf {\bibinfo {volume} {112}},\ \bibinfo {pages}
  {1900} (\bibinfo {year} {1958})}\BibitemShut {NoStop}%
\bibitem [{\citenamefont {Klimin}\ \emph {et~al.}(2019)\citenamefont {Klimin},
  \citenamefont {Kurkjian},\ and\ \citenamefont {Tempere}}]{Klimin2019}%
  \BibitemOpen
  \bibfield  {author} {\bibinfo {author} {\bibfnamefont {S.~N.}\ \bibnamefont
  {Klimin}}, \bibinfo {author} {\bibfnamefont {H.}~\bibnamefont {Kurkjian}},\
  and\ \bibinfo {author} {\bibfnamefont {J.}~\bibnamefont {Tempere}},\
  }\bibfield  {title} {\bibinfo {title} {Anderson-{B}ogoliubov collective
  excitations in superfluid {F}ermi gases at nonzero temperatures},\ }\href
  {https://doi.org/10.1007/s10909-019-02160-3} {\bibfield  {journal} {\bibinfo
  {journal} {J. Low Temp. Phys.}\ }\textbf {\bibinfo {volume} {196}},\ \bibinfo
  {pages} {102–110} (\bibinfo {year} {2019})}\BibitemShut {NoStop}%
\bibitem [{\citenamefont {Pascucci}\ \emph {et~al.}(2024)\citenamefont
  {Pascucci}, \citenamefont {Conti}, \citenamefont {Perali}, \citenamefont
  {Tempere},\ and\ \citenamefont {Neilson}}]{Pascucci2024}%
  \BibitemOpen
  \bibfield  {author} {\bibinfo {author} {\bibfnamefont {F.}~\bibnamefont
  {Pascucci}}, \bibinfo {author} {\bibfnamefont {S.}~\bibnamefont {Conti}},
  \bibinfo {author} {\bibfnamefont {A.}~\bibnamefont {Perali}}, \bibinfo
  {author} {\bibfnamefont {J.}~\bibnamefont {Tempere}},\ and\ \bibinfo {author}
  {\bibfnamefont {D.}~\bibnamefont {Neilson}},\ }\bibfield  {title} {\bibinfo
  {title} {Effects of intralayer correlations on electron-hole double-layer
  superfluidity},\ }\href {https://doi.org/10.1103/PhysRevB.109.094512}
  {\bibfield  {journal} {\bibinfo  {journal} {Phys. Rev. B}\ }\textbf {\bibinfo
  {volume} {109}},\ \bibinfo {pages} {094512} (\bibinfo {year}
  {2024})}\BibitemShut {NoStop}%
\bibitem [{\citenamefont {{Stratonovich}}(1957)}]{Stratonovich1957}%
  \BibitemOpen
  \bibfield  {author} {\bibinfo {author} {\bibfnamefont {R.~L.}\ \bibnamefont
  {{Stratonovich}}},\ }\bibfield  {title} {\bibinfo {title} {{On a method of
  calculating quantum distribution functions}},\ }\href@noop {} {\bibfield
  {journal} {\bibinfo  {journal} {Soviet Physics Doklady}\ }\textbf {\bibinfo
  {volume} {2}},\ \bibinfo {pages} {416} (\bibinfo {year} {1957})}\BibitemShut
  {NoStop}%
\bibitem [{\citenamefont {Ando}\ \emph {et~al.}(1982)\citenamefont {Ando},
  \citenamefont {Fowler},\ and\ \citenamefont {Stern}}]{Ando1982}%
  \BibitemOpen
  \bibfield  {author} {\bibinfo {author} {\bibfnamefont {T.}~\bibnamefont
  {Ando}}, \bibinfo {author} {\bibfnamefont {A.~B.}\ \bibnamefont {Fowler}},\
  and\ \bibinfo {author} {\bibfnamefont {F.}~\bibnamefont {Stern}},\ }\bibfield
   {title} {\bibinfo {title} {Electronic properties of two-dimensional
  systems},\ }\href {https://doi.org/10.1103/RevModPhys.54.437} {\bibfield
  {journal} {\bibinfo  {journal} {Rev. Mod. Phys.}\ }\textbf {\bibinfo {volume}
  {54}},\ \bibinfo {pages} {437} (\bibinfo {year} {1982})}\BibitemShut
  {NoStop}%
\bibitem [{\citenamefont {Scammell}\ and\ \citenamefont
  {Sushkov}(2023)}]{Scammel2023}%
  \BibitemOpen
  \bibfield  {author} {\bibinfo {author} {\bibfnamefont {H.~D.}\ \bibnamefont
  {Scammell}}\ and\ \bibinfo {author} {\bibfnamefont {O.~P.}\ \bibnamefont
  {Sushkov}},\ }\bibfield  {title} {\bibinfo {title} {Exciton condensation in
  biased bilayer graphene},\ }\href
  {https://doi.org/10.1103/PhysRevResearch.5.043176} {\bibfield  {journal}
  {\bibinfo  {journal} {Phys. Rev. Res.}\ }\textbf {\bibinfo {volume} {5}},\
  \bibinfo {pages} {043176} (\bibinfo {year} {2023})}\BibitemShut {NoStop}%
\bibitem [{\citenamefont {Conti}\ \emph {et~al.}(2017)\citenamefont {Conti},
  \citenamefont {Perali}, \citenamefont {Peeters},\ and\ \citenamefont
  {Neilson}}]{Conti2017}%
  \BibitemOpen
  \bibfield  {author} {\bibinfo {author} {\bibfnamefont {S.}~\bibnamefont
  {Conti}}, \bibinfo {author} {\bibfnamefont {A.}~\bibnamefont {Perali}},
  \bibinfo {author} {\bibfnamefont {F.~M.}\ \bibnamefont {Peeters}},\ and\
  \bibinfo {author} {\bibfnamefont {D.}~\bibnamefont {Neilson}},\ }\bibfield
  {title} {\bibinfo {title} {Multicomponent electron-hole superfluidity and the
  {BCS-BEC} crossover in double bilayer graphene},\ }\href
  {https://doi.org/10.1103/PhysRevLett.119.257002} {\bibfield  {journal}
  {\bibinfo  {journal} {Phys. Rev. Lett.}\ }\textbf {\bibinfo {volume} {119}},\
  \bibinfo {pages} {257002} (\bibinfo {year} {2017})}\BibitemShut {NoStop}%
\bibitem [{\citenamefont {H\o{}jlund}\ \emph {et~al.}(2023)\citenamefont
  {H\o{}jlund}, \citenamefont {Grovn}, \citenamefont {Pakdel}, \citenamefont
  {Thygesen},\ and\ \citenamefont {Nilsson}}]{Hojlund2023}%
  \BibitemOpen
  \bibfield  {author} {\bibinfo {author} {\bibfnamefont {R.}~\bibnamefont
  {H\o{}jlund}}, \bibinfo {author} {\bibfnamefont {E.}~\bibnamefont {Grovn}},
  \bibinfo {author} {\bibfnamefont {S.}~\bibnamefont {Pakdel}}, \bibinfo
  {author} {\bibfnamefont {K.~S.}\ \bibnamefont {Thygesen}},\ and\ \bibinfo
  {author} {\bibfnamefont {F.}~\bibnamefont {Nilsson}},\ }\bibfield  {title}
  {\bibinfo {title} {Exciton superfluidity in two-dimensional heterostructures
  from first principles: Importance of material-specific screening},\ }\href
  {https://doi.org/10.1103/PhysRevB.108.014506} {\bibfield  {journal} {\bibinfo
   {journal} {Phys. Rev. B}\ }\textbf {\bibinfo {volume} {108}},\ \bibinfo
  {pages} {014506} (\bibinfo {year} {2023})}\BibitemShut {NoStop}%
\bibitem [{\citenamefont {Devreese}\ \emph {et~al.}(2022)\citenamefont
  {Devreese}, \citenamefont {Tempere},\ and\ \citenamefont {S\'a~de
  Melo}}]{Devreese2022}%
  \BibitemOpen
  \bibfield  {author} {\bibinfo {author} {\bibfnamefont {J.~P.~A.}\
  \bibnamefont {Devreese}}, \bibinfo {author} {\bibfnamefont {J.}~\bibnamefont
  {Tempere}},\ and\ \bibinfo {author} {\bibfnamefont {C.~A.~R.}\ \bibnamefont
  {S\'a~de Melo}},\ }\bibfield  {title} {\bibinfo {title} {Topological phases
  and collective modes in {U}(1) and {SU}(2) sectors of spin-orbit-coupled
  two-dimensional superfluid fermi gases},\ }\href
  {https://doi.org/10.1103/PhysRevA.105.033304} {\bibfield  {journal} {\bibinfo
   {journal} {Phys. Rev. A}\ }\textbf {\bibinfo {volume} {105}},\ \bibinfo
  {pages} {033304} (\bibinfo {year} {2022})}\BibitemShut {NoStop}%
\bibitem [{\citenamefont {Benfatto}\ \emph {et~al.}(2004)\citenamefont
  {Benfatto}, \citenamefont {Toschi},\ and\ \citenamefont
  {Caprara}}]{Benfatto2004}%
  \BibitemOpen
  \bibfield  {author} {\bibinfo {author} {\bibfnamefont {L.}~\bibnamefont
  {Benfatto}}, \bibinfo {author} {\bibfnamefont {A.}~\bibnamefont {Toschi}},\
  and\ \bibinfo {author} {\bibfnamefont {S.}~\bibnamefont {Caprara}},\
  }\bibfield  {title} {\bibinfo {title} {Low-energy phase-only action in a
  superconductor: A comparison with the $\mathrm{XY}$ model},\ }\href
  {https://doi.org/10.1103/PhysRevB.69.184510} {\bibfield  {journal} {\bibinfo
  {journal} {Phys. Rev. B}\ }\textbf {\bibinfo {volume} {69}},\ \bibinfo
  {pages} {184510} (\bibinfo {year} {2004})}\BibitemShut {NoStop}%
\bibitem [{\citenamefont {Tempere}\ \emph {et~al.}(2009)\citenamefont
  {Tempere}, \citenamefont {Klimin},\ and\ \citenamefont
  {Devreese}}]{Tempere2009}%
  \BibitemOpen
  \bibfield  {author} {\bibinfo {author} {\bibfnamefont {J.}~\bibnamefont
  {Tempere}}, \bibinfo {author} {\bibfnamefont {S.~N.}\ \bibnamefont
  {Klimin}},\ and\ \bibinfo {author} {\bibfnamefont {J.~T.}\ \bibnamefont
  {Devreese}},\ }\bibfield  {title} {\bibinfo {title} {Effect of population
  imbalance on the berezinskii-kosterlitz-thouless phase transition in a
  superfluid fermi gas},\ }\href {https://doi.org/10.1103/PhysRevA.79.053637}
  {\bibfield  {journal} {\bibinfo  {journal} {Phys. Rev. A}\ }\textbf {\bibinfo
  {volume} {79}},\ \bibinfo {pages} {053637} (\bibinfo {year}
  {2009})}\BibitemShut {NoStop}%
\bibitem [{\citenamefont {Shi}\ \emph {et~al.}(2024)\citenamefont {Shi},
  \citenamefont {Zhang},\ and\ \citenamefont {Sá~de Melo}}]{Shi2024}%
  \BibitemOpen
  \bibfield  {author} {\bibinfo {author} {\bibfnamefont {T.}~\bibnamefont
  {Shi}}, \bibinfo {author} {\bibfnamefont {W.}~\bibnamefont {Zhang}},\ and\
  \bibinfo {author} {\bibfnamefont {C.~A.~R.}\ \bibnamefont {Sá~de Melo}},\
  }\bibfield  {title} {\bibinfo {title} {Tighter upper bounds on the critical
  temperature of two-dimensional superfluids and superconductors from the bcs
  to the bose regime},\ }\href {https://doi.org/10.1088/1367-2630/ad7281}
  {\bibfield  {journal} {\bibinfo  {journal} {New Journal of Physics}\ }\textbf
  {\bibinfo {volume} {26}},\ \bibinfo {pages} {093001} (\bibinfo {year}
  {2024})}\BibitemShut {NoStop}%
\bibitem [{\citenamefont {Botelho}\ and\ \citenamefont {S\'a~de
  Melo}(2006)}]{Botelho2006}%
  \BibitemOpen
  \bibfield  {author} {\bibinfo {author} {\bibfnamefont {S.~S.}\ \bibnamefont
  {Botelho}}\ and\ \bibinfo {author} {\bibfnamefont {C.~A.~R.}\ \bibnamefont
  {S\'a~de Melo}},\ }\bibfield  {title} {\bibinfo {title} {Vortex-antivortex
  lattice in ultracold fermionic gases},\ }\href
  {https://doi.org/10.1103/PhysRevLett.96.040404} {\bibfield  {journal}
  {\bibinfo  {journal} {Phys. Rev. Lett.}\ }\textbf {\bibinfo {volume} {96}},\
  \bibinfo {pages} {040404} (\bibinfo {year} {2006})}\BibitemShut {NoStop}%
\bibitem [{\citenamefont {Neilson}\ \emph {et~al.}(2014)\citenamefont
  {Neilson}, \citenamefont {Perali},\ and\ \citenamefont
  {Hamilton}}]{Neilson2014}%
  \BibitemOpen
  \bibfield  {author} {\bibinfo {author} {\bibfnamefont {D.}~\bibnamefont
  {Neilson}}, \bibinfo {author} {\bibfnamefont {A.}~\bibnamefont {Perali}},\
  and\ \bibinfo {author} {\bibfnamefont {A.~R.}\ \bibnamefont {Hamilton}},\
  }\bibfield  {title} {\bibinfo {title} {Excitonic superfluidity and screening
  in electron-hole bilayer systems},\ }\href
  {https://doi.org/10.1103/PhysRevB.89.060502} {\bibfield  {journal} {\bibinfo
  {journal} {Phys. Rev. B}\ }\textbf {\bibinfo {volume} {89}},\ \bibinfo
  {pages} {060502(R)} (\bibinfo {year} {2014})}\BibitemShut {NoStop}%
\bibitem [{\citenamefont {Benfatto}\ \emph {et~al.}(2007)\citenamefont
  {Benfatto}, \citenamefont {Castellani},\ and\ \citenamefont
  {Giamarchi}}]{Benfatto2007}%
  \BibitemOpen
  \bibfield  {author} {\bibinfo {author} {\bibfnamefont {L.}~\bibnamefont
  {Benfatto}}, \bibinfo {author} {\bibfnamefont {C.}~\bibnamefont
  {Castellani}},\ and\ \bibinfo {author} {\bibfnamefont {T.}~\bibnamefont
  {Giamarchi}},\ }\bibfield  {title} {\bibinfo {title} {Kosterlitz-thouless
  behavior in layered superconductors: The role of the vortex core energy},\
  }\href {https://doi.org/10.1103/PhysRevLett.98.117008} {\bibfield  {journal}
  {\bibinfo  {journal} {Phys. Rev. Lett.}\ }\textbf {\bibinfo {volume} {98}},\
  \bibinfo {pages} {117008} (\bibinfo {year} {2007})}\BibitemShut {NoStop}%
\bibitem [{\citenamefont {Castro}\ \emph {et~al.}(2007)\citenamefont {Castro},
  \citenamefont {Novoselov}, \citenamefont {Morozov}, \citenamefont {Peres},
  \citenamefont {dos Santos}, \citenamefont {Nilsson}, \citenamefont {Guinea},
  \citenamefont {Geim},\ and\ \citenamefont {Castro~Neto}}]{Castro2007}%
  \BibitemOpen
  \bibfield  {author} {\bibinfo {author} {\bibfnamefont {E.~V.}\ \bibnamefont
  {Castro}}, \bibinfo {author} {\bibfnamefont {K.~S.}\ \bibnamefont
  {Novoselov}}, \bibinfo {author} {\bibfnamefont {S.~V.}\ \bibnamefont
  {Morozov}}, \bibinfo {author} {\bibfnamefont {N.~M.~R.}\ \bibnamefont
  {Peres}}, \bibinfo {author} {\bibfnamefont {J.~M. B.~L.}\ \bibnamefont {dos
  Santos}}, \bibinfo {author} {\bibfnamefont {J.}~\bibnamefont {Nilsson}},
  \bibinfo {author} {\bibfnamefont {F.}~\bibnamefont {Guinea}}, \bibinfo
  {author} {\bibfnamefont {A.~K.}\ \bibnamefont {Geim}},\ and\ \bibinfo
  {author} {\bibfnamefont {A.~H.}\ \bibnamefont {Castro~Neto}},\ }\bibfield
  {title} {\bibinfo {title} {Biased bilayer graphene: Semiconductor with a gap
  tunable by the electric field effect},\ }\href
  {https://doi.org/10.1103/PhysRevLett.99.216802} {\bibfield  {journal}
  {\bibinfo  {journal} {Phys. Rev. Lett.}\ }\textbf {\bibinfo {volume} {99}},\
  \bibinfo {pages} {216802} (\bibinfo {year} {2007})}\BibitemShut {NoStop}%
\bibitem [{\citenamefont {Zou}\ \emph {et~al.}(2011)\citenamefont {Zou},
  \citenamefont {Hong},\ and\ \citenamefont {Zhu}}]{Zou2011}%
  \BibitemOpen
  \bibfield  {author} {\bibinfo {author} {\bibfnamefont {K.}~\bibnamefont
  {Zou}}, \bibinfo {author} {\bibfnamefont {X.}~\bibnamefont {Hong}},\ and\
  \bibinfo {author} {\bibfnamefont {J.}~\bibnamefont {Zhu}},\ }\bibfield
  {title} {\bibinfo {title} {Effective mass of electrons and holes in bilayer
  graphene: Electron-hole asymmetry and electron-electron interaction},\ }\href
  {https://doi.org/10.1103/PhysRevB.84.085408} {\bibfield  {journal} {\bibinfo
  {journal} {Phys. Rev. B}\ }\textbf {\bibinfo {volume} {84}},\ \bibinfo
  {pages} {085408} (\bibinfo {year} {2011})}\BibitemShut {NoStop}%
\bibitem [{\citenamefont {Lozovik}\ \emph {et~al.}(2012)\citenamefont
  {Lozovik}, \citenamefont {Ogarkov},\ and\ \citenamefont
  {Sokolik}}]{Lozovik2012}%
  \BibitemOpen
  \bibfield  {author} {\bibinfo {author} {\bibfnamefont {Y.~E.}\ \bibnamefont
  {Lozovik}}, \bibinfo {author} {\bibfnamefont {S.~L.}\ \bibnamefont
  {Ogarkov}},\ and\ \bibinfo {author} {\bibfnamefont {A.~A.}\ \bibnamefont
  {Sokolik}},\ }\bibfield  {title} {\bibinfo {title} {Condensation of
  electron-hole pairs in a two-layer graphene system: {C}orrelation effects},\
  }\href {https://doi.org/10.1103/PhysRevB.86.045429} {\bibfield  {journal}
  {\bibinfo  {journal} {Phys. Rev. B}\ }\textbf {\bibinfo {volume} {86}},\
  \bibinfo {pages} {045429} (\bibinfo {year} {2012})}\BibitemShut {NoStop}%
\bibitem [{\citenamefont {Weitzel}\ \emph {et~al.}(2023)\citenamefont
  {Weitzel}, \citenamefont {Pfaffinger}, \citenamefont {Maccari}, \citenamefont
  {Kronfeldner}, \citenamefont {Huber}, \citenamefont {Fuchs}, \citenamefont
  {Mallord}, \citenamefont {Linzen}, \citenamefont {Il'ichev}, \citenamefont
  {Paradiso},\ and\ \citenamefont {Strunk}}]{Weitzel2023}%
  \BibitemOpen
  \bibfield  {author} {\bibinfo {author} {\bibfnamefont {A.}~\bibnamefont
  {Weitzel}}, \bibinfo {author} {\bibfnamefont {L.}~\bibnamefont {Pfaffinger}},
  \bibinfo {author} {\bibfnamefont {I.}~\bibnamefont {Maccari}}, \bibinfo
  {author} {\bibfnamefont {K.}~\bibnamefont {Kronfeldner}}, \bibinfo {author}
  {\bibfnamefont {T.}~\bibnamefont {Huber}}, \bibinfo {author} {\bibfnamefont
  {L.}~\bibnamefont {Fuchs}}, \bibinfo {author} {\bibfnamefont
  {J.}~\bibnamefont {Mallord}}, \bibinfo {author} {\bibfnamefont
  {S.}~\bibnamefont {Linzen}}, \bibinfo {author} {\bibfnamefont
  {E.}~\bibnamefont {Il'ichev}}, \bibinfo {author} {\bibfnamefont
  {N.}~\bibnamefont {Paradiso}},\ and\ \bibinfo {author} {\bibfnamefont
  {C.}~\bibnamefont {Strunk}},\ }\bibfield  {title} {\bibinfo {title}
  {Sharpness of the berezinskii-kosterlitz-thouless transition in disordered
  nbn films},\ }\href {https://doi.org/10.1103/PhysRevLett.131.186002}
  {\bibfield  {journal} {\bibinfo  {journal} {Phys. Rev. Lett.}\ }\textbf
  {\bibinfo {volume} {131}},\ \bibinfo {pages} {186002} (\bibinfo {year}
  {2023})}\BibitemShut {NoStop}%
\bibitem [{\citenamefont {Midei}\ \emph {et~al.}(2024)\citenamefont {Midei},
  \citenamefont {Furutani}, \citenamefont {Salasnich},\ and\ \citenamefont
  {Perali}}]{Midei2024}%
  \BibitemOpen
  \bibfield  {author} {\bibinfo {author} {\bibfnamefont {G.}~\bibnamefont
  {Midei}}, \bibinfo {author} {\bibfnamefont {K.}~\bibnamefont {Furutani}},
  \bibinfo {author} {\bibfnamefont {L.}~\bibnamefont {Salasnich}},\ and\
  \bibinfo {author} {\bibfnamefont {A.}~\bibnamefont {Perali}},\ }\bibfield
  {title} {\bibinfo {title} {Predictive power of the
  berezinskii-kosterlitz-thouless theory based on renormalization group
  throughout the bcs-bec crossover in two-dimensional superconductors},\ }\href
  {https://doi.org/10.1103/PhysRevB.110.214502} {\bibfield  {journal} {\bibinfo
   {journal} {Phys. Rev. B}\ }\textbf {\bibinfo {volume} {110}},\ \bibinfo
  {pages} {214502} (\bibinfo {year} {2024})}\BibitemShut {NoStop}%
\bibitem [{\citenamefont {Emery}\ and\ \citenamefont
  {Kivelson}(1995)}]{Emery1995}%
  \BibitemOpen
  \bibfield  {author} {\bibinfo {author} {\bibfnamefont {V.~J.}\ \bibnamefont
  {Emery}}\ and\ \bibinfo {author} {\bibfnamefont {S.~A.}\ \bibnamefont
  {Kivelson}},\ }\bibfield  {title} {\bibinfo {title} {Importance of phase
  fluctuations in superconductors with small superfluid density},\ }\href
  {https://doi.org/10.1038/374434a0} {\bibfield  {journal} {\bibinfo  {journal}
  {Nature}\ }\textbf {\bibinfo {volume} {374}},\ \bibinfo {pages} {434}
  (\bibinfo {year} {1995})}\BibitemShut {NoStop}%
\bibitem [{\citenamefont {Timusk}\ and\ \citenamefont
  {Statt}(1999)}]{Timusk1999}%
  \BibitemOpen
  \bibfield  {author} {\bibinfo {author} {\bibfnamefont {T.}~\bibnamefont
  {Timusk}}\ and\ \bibinfo {author} {\bibfnamefont {B.}~\bibnamefont {Statt}},\
  }\bibfield  {title} {\bibinfo {title} {The pseudogap in high-temperature
  superconductors: an experimental survey},\ }\href
  {https://doi.org/10.1088/0034-4885/62/1/002} {\bibfield  {journal} {\bibinfo
  {journal} {Reports on Progress in Physics}\ }\textbf {\bibinfo {volume}
  {62}},\ \bibinfo {pages} {61} (\bibinfo {year} {1999})}\BibitemShut {NoStop}%
\bibitem [{\citenamefont {Lee}\ \emph {et~al.}(2009)\citenamefont {Lee},
  \citenamefont {Fujita}, \citenamefont {Schmidt}, \citenamefont {Kim},
  \citenamefont {Eisaki}, \citenamefont {Uchida},\ and\ \citenamefont
  {Davis}}]{Lee2009}%
  \BibitemOpen
  \bibfield  {author} {\bibinfo {author} {\bibfnamefont {J.}~\bibnamefont
  {Lee}}, \bibinfo {author} {\bibfnamefont {K.}~\bibnamefont {Fujita}},
  \bibinfo {author} {\bibfnamefont {A.~R.}\ \bibnamefont {Schmidt}}, \bibinfo
  {author} {\bibfnamefont {C.~K.}\ \bibnamefont {Kim}}, \bibinfo {author}
  {\bibfnamefont {H.}~\bibnamefont {Eisaki}}, \bibinfo {author} {\bibfnamefont
  {S.}~\bibnamefont {Uchida}},\ and\ \bibinfo {author} {\bibfnamefont {J.~C.}\
  \bibnamefont {Davis}},\ }\bibfield  {title} {\bibinfo {title} {Spectroscopic
  fingerprint of phase-incoherent superconductivity in the underdoped
  {B}i$_2${S}r$_2${C}a{C}u$_2${O}$_8$$_+$$_\delta$},\ }\href
  {https://doi.org/10.1126/science.1176369} {\bibfield  {journal} {\bibinfo
  {journal} {Science}\ }\textbf {\bibinfo {volume} {325}},\ \bibinfo {pages}
  {1099} (\bibinfo {year} {2009})},\ \Eprint
  {https://arxiv.org/abs/https://www.science.org/doi/pdf/10.1126/science.1176369}
  {https://www.science.org/doi/pdf/10.1126/science.1176369} \BibitemShut
  {NoStop}%
\bibitem [{\citenamefont {Salasnich}\ and\ \citenamefont
  {Toigo}(2015)}]{Salasnich2015}%
  \BibitemOpen
  \bibfield  {author} {\bibinfo {author} {\bibfnamefont {L.}~\bibnamefont
  {Salasnich}}\ and\ \bibinfo {author} {\bibfnamefont {F.}~\bibnamefont
  {Toigo}},\ }\bibfield  {title} {\bibinfo {title} {Composite bosons in the
  two--dimensional {BCS--BEC} crossover from {G}aussian fluctuations},\ }\href
  {https://doi.org/10.1103/PhysRevA.91.011604} {\bibfield  {journal} {\bibinfo
  {journal} {Phys. Rev. A}\ }\textbf {\bibinfo {volume} {91}},\ \bibinfo
  {pages} {011604} (\bibinfo {year} {2015})}\BibitemShut {NoStop}%
\bibitem [{\citenamefont {Combescot}\ \emph {et~al.}(2006)\citenamefont
  {Combescot}, \citenamefont {Kagan},\ and\ \citenamefont
  {Stringari}}]{Combescot2006}%
  \BibitemOpen
  \bibfield  {author} {\bibinfo {author} {\bibfnamefont {R.}~\bibnamefont
  {Combescot}}, \bibinfo {author} {\bibfnamefont {M.~Y.}\ \bibnamefont
  {Kagan}},\ and\ \bibinfo {author} {\bibfnamefont {S.}~\bibnamefont
  {Stringari}},\ }\bibfield  {title} {\bibinfo {title} {Collective mode of
  homogeneous superfluid {Fermi} gases in the {BEC-BCS} crossover},\ }\href
  {https://doi.org/10.1103/PhysRevA.74.042717} {\bibfield  {journal} {\bibinfo
  {journal} {Phys. Rev. A}\ }\textbf {\bibinfo {volume} {74}},\ \bibinfo
  {pages} {042717} (\bibinfo {year} {2006})}\BibitemShut {NoStop}%
\bibitem [{\citenamefont {Kuhn}\ \emph {et~al.}(2020)\citenamefont {Kuhn},
  \citenamefont {Hoinka}, \citenamefont {Herrera}, \citenamefont {Dyke},
  \citenamefont {Kinnunen}, \citenamefont {Bruun},\ and\ \citenamefont
  {Vale}}]{Kuhn2020}%
  \BibitemOpen
  \bibfield  {author} {\bibinfo {author} {\bibfnamefont {C.~C.~N.}\
  \bibnamefont {Kuhn}}, \bibinfo {author} {\bibfnamefont {S.}~\bibnamefont
  {Hoinka}}, \bibinfo {author} {\bibfnamefont {I.}~\bibnamefont {Herrera}},
  \bibinfo {author} {\bibfnamefont {P.}~\bibnamefont {Dyke}}, \bibinfo {author}
  {\bibfnamefont {J.~J.}\ \bibnamefont {Kinnunen}}, \bibinfo {author}
  {\bibfnamefont {G.~M.}\ \bibnamefont {Bruun}},\ and\ \bibinfo {author}
  {\bibfnamefont {C.~J.}\ \bibnamefont {Vale}},\ }\bibfield  {title} {\bibinfo
  {title} {High-frequency sound in a unitary fermi gas},\ }\href
  {https://doi.org/10.1103/PhysRevLett.124.150401} {\bibfield  {journal}
  {\bibinfo  {journal} {Phys. Rev. Lett.}\ }\textbf {\bibinfo {volume} {124}},\
  \bibinfo {pages} {150401} (\bibinfo {year} {2020})}\BibitemShut {NoStop}%
\bibitem [{\citenamefont {Kurkjian}\ \emph {et~al.}(2019)\citenamefont
  {Kurkjian}, \citenamefont {Klimin}, \citenamefont {Tempere},\ and\
  \citenamefont {Castin}}]{Kurkjian2019}%
  \BibitemOpen
  \bibfield  {author} {\bibinfo {author} {\bibfnamefont {H.}~\bibnamefont
  {Kurkjian}}, \bibinfo {author} {\bibfnamefont {S.~N.}\ \bibnamefont
  {Klimin}}, \bibinfo {author} {\bibfnamefont {J.}~\bibnamefont {Tempere}},\
  and\ \bibinfo {author} {\bibfnamefont {Y.}~\bibnamefont {Castin}},\
  }\bibfield  {title} {\bibinfo {title} {Pair-breaking collective branch in
  {BCS} superconductors and superfluid {F}ermi gases},\ }\href
  {https://doi.org/10.1103/PhysRevLett.122.093403} {\bibfield  {journal}
  {\bibinfo  {journal} {Phys. Rev. Lett.}\ }\textbf {\bibinfo {volume} {122}},\
  \bibinfo {pages} {093403} (\bibinfo {year} {2019})}\BibitemShut {NoStop}%
\bibitem [{\citenamefont {Valtolina}\ \emph {et~al.}(2015)\citenamefont
  {Valtolina}, \citenamefont {Burchianti}, \citenamefont {Amico}, \citenamefont
  {Neri}, \citenamefont {Xhani}, \citenamefont {Seman}, \citenamefont
  {Trombettoni}, \citenamefont {Smerzi}, \citenamefont {Zaccanti},
  \citenamefont {Inguscio},\ and\ \citenamefont {Roati}}]{Valtolina2015}%
  \BibitemOpen
  \bibfield  {author} {\bibinfo {author} {\bibfnamefont {G.}~\bibnamefont
  {Valtolina}}, \bibinfo {author} {\bibfnamefont {A.}~\bibnamefont
  {Burchianti}}, \bibinfo {author} {\bibfnamefont {A.}~\bibnamefont {Amico}},
  \bibinfo {author} {\bibfnamefont {E.}~\bibnamefont {Neri}}, \bibinfo {author}
  {\bibfnamefont {K.}~\bibnamefont {Xhani}}, \bibinfo {author} {\bibfnamefont
  {J.~A.}\ \bibnamefont {Seman}}, \bibinfo {author} {\bibfnamefont
  {A.}~\bibnamefont {Trombettoni}}, \bibinfo {author} {\bibfnamefont
  {A.}~\bibnamefont {Smerzi}}, \bibinfo {author} {\bibfnamefont
  {M.}~\bibnamefont {Zaccanti}}, \bibinfo {author} {\bibfnamefont
  {M.}~\bibnamefont {Inguscio}},\ and\ \bibinfo {author} {\bibfnamefont
  {G.}~\bibnamefont {Roati}},\ }\bibfield  {title} {\bibinfo {title} {Josephson
  effect in fermionic superfluids across the bec-bcs crossover},\ }\href
  {https://doi.org/10.1126/science.aac9725} {\bibfield  {journal} {\bibinfo
  {journal} {Science}\ }\textbf {\bibinfo {volume} {350}},\ \bibinfo {pages}
  {1505} (\bibinfo {year} {2015})},\ \Eprint
  {https://arxiv.org/abs/https://www.science.org/doi/pdf/10.1126/science.aac9725}
  {https://www.science.org/doi/pdf/10.1126/science.aac9725} \BibitemShut
  {NoStop}%
\bibitem [{\citenamefont {Conti}\ \emph {et~al.}(2023)\citenamefont {Conti},
  \citenamefont {Perali}, \citenamefont {Hamilton}, \citenamefont {Milo\ifmmode
  \check{s}\else \v{s}\fi{}evi\ifmmode~\acute{c}\else \'{c}\fi{}},
  \citenamefont {Peeters},\ and\ \citenamefont {Neilson}}]{Conti2023}%
  \BibitemOpen
  \bibfield  {author} {\bibinfo {author} {\bibfnamefont {S.}~\bibnamefont
  {Conti}}, \bibinfo {author} {\bibfnamefont {A.}~\bibnamefont {Perali}},
  \bibinfo {author} {\bibfnamefont {A.~R.}\ \bibnamefont {Hamilton}}, \bibinfo
  {author} {\bibfnamefont {M.~V.}\ \bibnamefont {Milo\ifmmode \check{s}\else
  \v{s}\fi{}evi\ifmmode~\acute{c}\else \'{c}\fi{}}}, \bibinfo {author}
  {\bibfnamefont {F.~M.}\ \bibnamefont {Peeters}},\ and\ \bibinfo {author}
  {\bibfnamefont {D.}~\bibnamefont {Neilson}},\ }\bibfield  {title} {\bibinfo
  {title} {Chester supersolid of spatially indirect excitons in double-layer
  semiconductor heterostructures},\ }\href
  {https://doi.org/10.1103/PhysRevLett.130.057001} {\bibfield  {journal}
  {\bibinfo  {journal} {Phys. Rev. Lett.}\ }\textbf {\bibinfo {volume} {130}},\
  \bibinfo {pages} {057001} (\bibinfo {year} {2023})}\BibitemShut {NoStop}%
\bibitem [{\citenamefont {Furutani}\ \emph {et~al.}(2024)\citenamefont
  {Furutani}, \citenamefont {Midei}, \citenamefont {Perali},\ and\
  \citenamefont {Salasnich}}]{Furutani2024}%
  \BibitemOpen
  \bibfield  {author} {\bibinfo {author} {\bibfnamefont {K.}~\bibnamefont
  {Furutani}}, \bibinfo {author} {\bibfnamefont {G.}~\bibnamefont {Midei}},
  \bibinfo {author} {\bibfnamefont {A.}~\bibnamefont {Perali}},\ and\ \bibinfo
  {author} {\bibfnamefont {L.}~\bibnamefont {Salasnich}},\ }\bibfield  {title}
  {\bibinfo {title} {Amplitude, phase, and topological fluctuations shaping the
  complex phase diagram of two-dimensional superconductors},\ }\href
  {https://doi.org/10.1103/PhysRevB.110.134501} {\bibfield  {journal} {\bibinfo
   {journal} {Phys. Rev. B}\ }\textbf {\bibinfo {volume} {110}},\ \bibinfo
  {pages} {134501} (\bibinfo {year} {2024})}\BibitemShut {NoStop}%
\bibitem [{\citenamefont {Conti}\ \emph {et~al.}(2025)\citenamefont {Conti},
  \citenamefont {Chaves}, \citenamefont {Ardila}, \citenamefont {Neilson},\
  and\ \citenamefont {Milo{\v{s}}evi{\'c}}}]{Conti2025}%
  \BibitemOpen
  \bibfield  {author} {\bibinfo {author} {\bibfnamefont {S.}~\bibnamefont
  {Conti}}, \bibinfo {author} {\bibfnamefont {A.}~\bibnamefont {Chaves}},
  \bibinfo {author} {\bibfnamefont {L.~A.~P.}\ \bibnamefont {Ardila}}, \bibinfo
  {author} {\bibfnamefont {D.}~\bibnamefont {Neilson}},\ and\ \bibinfo {author}
  {\bibfnamefont {M.~V.}\ \bibnamefont {Milo{\v{s}}evi{\'c}}},\ }\bibfield
  {title} {\bibinfo {title} {Vortices in dipolar condensates of interlayer
  excitons},\ }\href {https://doi.org/10.1103/sjmx-58ny} {\bibfield  {journal}
  {\bibinfo  {journal} {Phys. Rev. B}\ }\textbf {\bibinfo {volume} {112}},\
  \bibinfo {pages} {184514} (\bibinfo {year} {2025})}\BibitemShut {NoStop}%
\bibitem [{\citenamefont {Haas}\ and\ \citenamefont
  {Eliasson}(2018)}]{Haas2018}%
  \BibitemOpen
  \bibfield  {author} {\bibinfo {author} {\bibfnamefont {F.}~\bibnamefont
  {Haas}}\ and\ \bibinfo {author} {\bibfnamefont {B.}~\bibnamefont
  {Eliasson}},\ }\bibfield  {title} {\bibinfo {title} {Time-dependent
  variational approach for {B}ose--{E}instein condensates with nonlocal
  interaction},\ }\href {https://doi.org/10.1088/1361-6455/aad629} {\bibfield
  {journal} {\bibinfo  {journal} {J. Phys. B: At. Mol. Opt. Phys.}\ }\textbf
  {\bibinfo {volume} {51}},\ \bibinfo {pages} {175302} (\bibinfo {year}
  {2018})}\BibitemShut {NoStop}%
\bibitem [{\citenamefont {Fischer}(2006)}]{Fischer2006}%
  \BibitemOpen
  \bibfield  {author} {\bibinfo {author} {\bibfnamefont {U.~R.}\ \bibnamefont
  {Fischer}},\ }\bibfield  {title} {\bibinfo {title} {Stability of
  quasi-two-dimensional {Bose-Einstein} condensates with dominant dipole-dipole
  interactions},\ }\href {https://doi.org/10.1103/PhysRevA.73.031602}
  {\bibfield  {journal} {\bibinfo  {journal} {Phys. Rev. A}\ }\textbf {\bibinfo
  {volume} {73}},\ \bibinfo {pages} {031602} (\bibinfo {year}
  {2006})}\BibitemShut {NoStop}%
\bibitem [{\citenamefont {Al~Khawaja}\ \emph {et~al.}(2002)\citenamefont
  {Al~Khawaja}, \citenamefont {Andersen}, \citenamefont {Proukakis},\ and\
  \citenamefont {Stoof}}]{AlKhawaja2002}%
  \BibitemOpen
  \bibfield  {author} {\bibinfo {author} {\bibfnamefont {U.}~\bibnamefont
  {Al~Khawaja}}, \bibinfo {author} {\bibfnamefont {J.~O.}\ \bibnamefont
  {Andersen}}, \bibinfo {author} {\bibfnamefont {N.~P.}\ \bibnamefont
  {Proukakis}},\ and\ \bibinfo {author} {\bibfnamefont {H.~T.~C.}\ \bibnamefont
  {Stoof}},\ }\bibfield  {title} {\bibinfo {title} {Low dimensional bose
  gases},\ }\href {https://doi.org/10.1103/PhysRevA.66.013615} {\bibfield
  {journal} {\bibinfo  {journal} {Phys. Rev. A}\ }\textbf {\bibinfo {volume}
  {66}},\ \bibinfo {pages} {013615} (\bibinfo {year} {2002})}\BibitemShut
  {NoStop}%
\bibitem [{\citenamefont {Pascucci}(2024)}]{Pascucci2024thesis}%
  \BibitemOpen
  \bibfield  {author} {\bibinfo {author} {\bibfnamefont {F.}~\bibnamefont
  {Pascucci}},\ }\emph {\bibinfo {title} {Superfluidity in Exciton Bilayer
  Systems: Josephson Effect and Collective Modes as Definitive
  Identification-Markers}},\ \href
  {https://doi.org/10.63028/10067/2078520151162165141} {Ph.D. thesis},\
  \bibinfo  {school} {Universit{\`a} degli Studi di Camerino and Universiteit
  Antwerpen}, \bibinfo {address} {Place of publication not known} (\bibinfo
  {year} {2024}),\ \bibinfo {note} {doctoral thesis. Promoters: Andrea Perali,
  Jacques Tempere, and David Neilson}\BibitemShut {NoStop}%
\end{thebibliography}
%

\appendix
\clearpage

\section{End Matter}
\textit{
Appendix 1: Finite temperature RPA screened interaction}\\

We include the superfluid screening in the gap equation, Eq.~\eqref{gapeq}, by evaluating the screened attractive electron-hole interaction $V_{eh}^{\mathrm{RPA}}$ within the self-consistent finite-temperature RPA~\cite{Neilson2014},
\begin{align}
 V^{RPA}_{eh}(\mathbf{q},T)\!=\!\frac{V^D_\mathbf{q}+\Pi^A_\mathbf{q}(T)\mathcal{A}_\mathbf{q}}{1\!-\!2(\Pi^N_\mathbf{q}(T)V^{S}_\mathbf{q}+\Pi^A_\mathbf{q}(T)V^{D}_\mathbf{q})+\mathcal{B}_{\mathbf{q}}(T)\mathcal{A}_\mathbf{q}}
 \label{eq-due}
\end{align}
with 
\begin{equation}
\mathcal{A}_\mathbf{q}\!=\!\left(V^{S}_\mathbf{q}\right)^2-\left(V^{D}_\mathbf{q}\right)^2, \,\,\mathcal{B}_{\mathbf{q}}(T)\!=\!\left(\Pi^N_\mathbf{q}(T)\right)^2-\left(\Pi^A_\mathbf{q}(T)\right)^2. \label{eq-tre}
\end{equation}
$\Pi_\mathbf{q}^N(T)$ and $\Pi_\mathbf{q}^A(T)$ are the static normal and anomalous polarization functions at temperature $T$,
\begin{align}
\Pi_{\mathbf{q}}^N(T) &\!=\!\sum_k
(u_{\mathbf{k}}^2u_{\mathbf{k+q}}^2 \!+\! v_{\mathbf{q}}^2v_{\mathbf{k+q}}^2)\frac{n_F(E_\mathbf{k},T)\!-\!n_F(E_{\mathbf{k+q}},T)}
{E_\mathbf{k} - E_{\mathbf{k+q}}}\nonumber\\
&\!-(v_{\mathbf{k}}^2u_{\mathbf{k+q}}^2 \!+\! u_{\mathbf{k}}^2v_{\mathbf{k+q}}^2)\frac{1\!-\!n_F(E_\mathbf{k},T)\!-\!n_F(E_{\mathbf{k+q}},T)}
{E_\mathbf{k} + E_{\mathbf{k+q}}} \label{PiNT}\\
\Pi_\mathbf{q}^A(T) &= -2u_\mathbf{k}v_\mathbf{k}u_{\mathbf{k+q}}v_{\mathbf{k+q}}
\Bigg[\frac{n_F(E_\mathbf{k},T)-n_F(E_{\mathbf{k+q}},T)}{E_\mathbf{k} - E_{\mathbf{k+q}}}\nonumber\\
&\qquad+\frac{1-n_F(E_\mathbf{k},T)-n_F(E_{\mathbf{k+q}},T)}{E_\mathbf{k}+E_{\mathbf{k+q}}}\Bigg]\ , \label{PiAT}
\end{align}
where $v_\mathbf{k}^2=1/2(1-\epsilon_\mathbf{k}/E_\mathbf{k})$ and $u_\mathbf{k}^2=1/2(1+\epsilon_\mathbf{k}/E_\mathbf{k})$ are the Bogoliubov amplitudes. We use the abbreviated notation, $E_\mathbf{k}(T)\rightarrow E_\mathbf{k}$. 
The bare Coulomb interactions in the effective mass approximation for 2D bilayers of electrons (e) and holes (h), $V^{S}_{\mathbf{q}}=V^{ee}_{\mathbf{q}}= V^{hh}_{\mathbf{q}}= \frac{2\pi e^2}{4\pi\varepsilon |\mathbf{q}|}$ and $ V^D_{\mathbf{q}}=V^{eh}_{\mathbf{q}}=V^{he}_{\mathbf{q}}=-V^S_{\mathbf{q}}\textrm{e}^{-|\mathbf{q}|d}$ \cite{Ando1982}. In Eqs.~\eqref{PiNT}-\eqref{PiAT}, the first term comes from pair-breaking due to thermal excitations. It vanishes at $T=0$.
\\
\\
\textit{Appendix 2: Exciton bilayer vortex-core energy}\\

The vortex energy is the energy systems difference $\Delta E \equiv E[\psi_v]-E[\psi_0]$, where $\psi_0=\sqrt{n_0}$ is the uniform wave-function of the superfluid at fixed particle number density $n_0$ and $\psi_v$ is the wave-function with one vortex.
We recall for a contact interaction of coupling strength $g$, the Gross-Pitaevskij (GP) energy functional $E[\psi]$ is,
\begin{equation}
E[\psi] = \int d^2\mathbf{r}\left[\frac{1}{2m_X^*}|\nabla\psi|^2
+ \frac{g}{2}|\psi|^4\right].
\label{eq:GP_functional}
\end{equation}
For $\psi_0$, the kinetic term vanishes and $E[\psi_0]=\tfrac{g}{2}n_0^2 A$, with $A$ the system area. For $\psi_v$, the linear term in $(|\psi_v|^2-n_0)$ cancels from particle-number conservation. So,
\begin{equation}
\Delta E = \int d^2\mathbf{r}\left[\frac{1}{2m_X^*}|\nabla\psi_v|^2
+ \frac{g}{2}\left(|\psi_v|^2 - n_0\right)^2\right].
\label{eq:DeltaE_canonical}
\end{equation}
For a singly quantized vortex centered at the origin, the ansatz
\begin{equation}
\psi_v(\mathbf{r}) = \sqrt{n_0}\,f(r/\xi)\,e^{i\vartheta},
\qquad
\xi^2 \equiv \frac{1}{2 m_X^* g n_0}
\label{eq:vortex_ansatz}
\end{equation}
is adopted, where $\vartheta$ is the azimuthal angle, $f(0)=0$ and $f(x)\to1$ as
$x\to\infty$. Inserting Eq.~\eqref{eq:vortex_ansatz} into
Eq.~\eqref{eq:DeltaE_canonical}, and performing the angular integration with a variable change $x=r/\xi$, the vortex-core energy $E_c$ is obtained by discarding the centrifugal term $f^2/x^2$,
\begin{equation}
E_c = \frac{1}{2m_X^*}\,n_0\,\pi
\int_0^\infty dx\,x\left[ 2\left(\frac{df}{dx}\right)^2+\left(1 - f^2\right)^2
\right].
\label{eq:core_energy}
\end{equation}

We have extended the formalism to the  exciton bilayer with non-contact interactions. The exciton GP energy functional is,
\begin{equation}
\begin{split}
E_X[\psi] = {} & \int d^2\mathbf{r}\,\frac{1}{2m_X^*}|\nabla\psi|^2 \\
           & + \frac{1}{2}\int d^2\mathbf{r}\,d^2\mathbf{r}'\,
             |\psi(\mathbf{r})|^2\,V_{XX}(|\mathbf{r}-\mathbf{r}'|)\,
             |\psi(\mathbf{r}')|^2 \ . 
\end{split}
\label{eq:GP_nonlocal}
\end{equation}
where $V_{XX}(\mathbf{r})$ is the exciton-exciton interaction~\cite{Conti2025},
\begin{equation}
V_{XX}(\mathbf{r}) = \frac{2e^2}{4\pi\epsilon}\left(\frac{1}{\mathbf{r}} - \frac{1}{\sqrt{\mathbf{r}^2 + d^2}}\right) \ .
\label{eq:VXX}
\end{equation}
$\mathbf{r}$ is the in-plane interparticle vector and $d$ the layer separation. For $d\ll r_0$, the  low density regime,  $V_{XX}$ is a dipolar interaction. The dipole moment is tunable through layer separation $d$. 

The kinetic term in Eq.\ \eqref{eq:GP_nonlocal} is the same as for the contact interaction, but the interaction is a nonlocal double integral with kernel $V_{XX}$. 
We compared the vortex profile and core energy for contact and exciton-exciton interactions. For the contact interaction we chose an effective contact coupling $g_{\mathrm{eff}}$ defined as the zero-momentum limit of the dipole-dipole interaction,  
\begin{equation}
V_{\mathrm{dd}}(k)=8\pi\,\frac{1-e^{-dk}}{k}\ ,
\end{equation}
\begin{equation}
g_{\mathrm{eff}}=\lim_{k\to0}V_{\mathrm{dd}}(k)=8\pi d.
\end{equation}
The units are [Ry$^*$a$_B^{*^2}$].
This generates the contact interaction that reproduces the dipolar kernel in the long-wavelength limit.

For the DBG parameters used in the main text, and with $d=2\,\mathrm{nm}\simeq0.253\,a_B^*$, this gives $g_{\mathrm{eff}}\simeq6.36\,\mathrm{Ry}^*(a_B^*)^2$. 
We use this value of $g_{\mathrm{eff}}$ to evaluate the local functional  (Eq.~\eqref{eq:GP_functional}) and  $V_{XX}$ (Eq.~\eqref{eq:VXX}) to evaluate the nonlocal functional, Eq.~\eqref{eq:GP_nonlocal}.
For the wave functions in Eqs.~\eqref{eq:GP_functional} and~\eqref{eq:GP_nonlocal}, we solved the GP equations for the contact interaction, 
\begin{equation}
-\frac{1}{2m_X^*}\nabla^2\Psi(\mathbf r)+g_{\mathrm{eff}}|\Psi(\mathbf r)|^2\Psi(\mathbf r)
= \mu \Psi(\mathbf r)\ ,
\label{GPc}
\end{equation}
and for the dipolar interaction,
\begin{equation}
-\frac{1}{2m_X^*}\nabla^2 \Psi(\mathbf{r})
+ \phi_{XX}(\mathbf{r}) \Psi(\mathbf{r})
= \mu \Psi(\mathbf{r})\ ,
\label{GPd}
\end{equation}
%
where $\phi_{XX}$ is the exciton mean-field potential \cite{Conti2025},
\begin{equation}
\phi_{XX}(\mathbf{r}) =
\int V_{XX}(|\mathbf{r}-\mathbf{r}'|)\,|\Psi(\mathbf{r}')|^2\, d\mathbf{r}'.
\end{equation}

We solved Eqs.~\eqref{GPc} and~\eqref{GPd} by imaginary-time evolution~\cite{Conti2025, Haas2018}, starting from a randomized order parameter, and evaluating $\phi_{XX}$ in Fourier space through the convolution theorem~\cite{Fischer2006}. The simulations were performed on a $2000^2$ grid with Dirichlet boundary conditions. Convergence was reached when the relative energy change between successive iterations fell below $10^{-10}$.

\begin{figure}[t]
\includegraphics[width=0.5\textwidth]{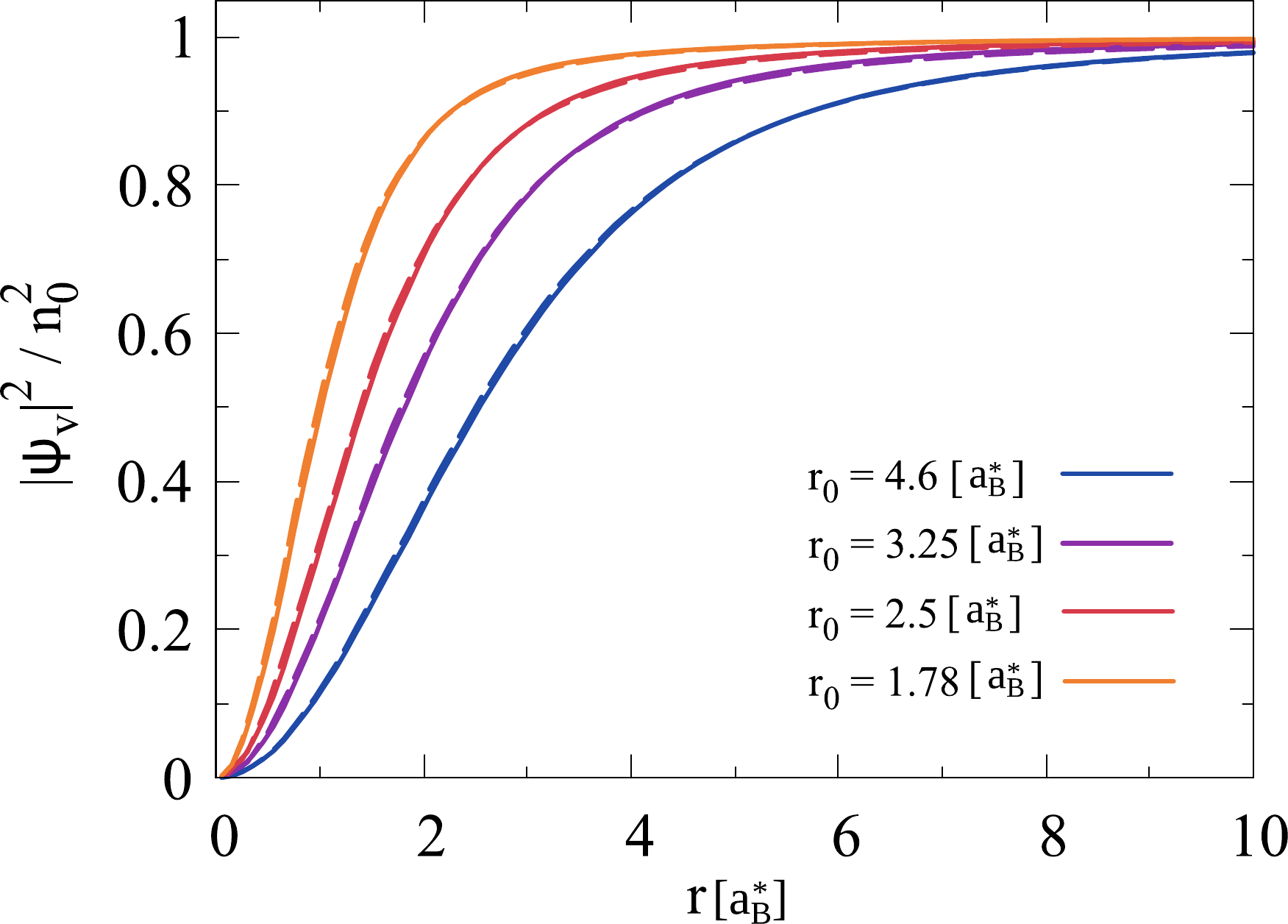}
\caption{Vortex density profiles $|\psi_v(r)|^2/n^2_0$ as a function of the radial distance $r$, obtained from the imaginary-time solutions of the GP equations for contact interactions (solid lines) and dipolar  interactions (dashed lines). The values of the inlayer interparticle distance $r_0$ span the density range addressed in this work.}
\label{vortex}
\end{figure}

The vortex density profiles obtained with the contact and dipolar interactions are shown in Fig.~\ref{vortex} for four densities. In the low density  regime which is our primary interest here, the two profiles are essentially indistinguishable. This reflects the fact that the excitons are compact and widely spaced which makes the vortex structure completely insensitive to whether the interaction is local or long-ranged.

For the contact interaction, the vortex-core energy is universal and density independent, with $E_c/J_0\simeq 2.45$ \cite{Shi2024, AlKhawaja2002}.
For the dipolar gas, we find a comparable value that remains almost constant over the density range studied. This is consistent with the vortex profiles in Fig.\ \ref{vortex}.
Deviations from the contact-interaction result can be expected at higher densities or at larger interlayer separations, where the details of the nonlocal form of the dipolar interaction become important. In this regime, a peak can develop near the vortex edge, signaling an accumulation of superfluid density around the core~\cite{Conti2025}.

\clearpage
\onecolumngrid
\section{Supplemental material: Spin-wave fluctuations}

We expand the action $S=S_0+S_{\mathrm{int}}(|\Delta|)+S_{\mathrm{ph}}$ around its saddle point by writing $|\Delta|=|\Delta_0|+\delta|\Delta|$ and neglecting amplitude fluctuations. The
action then decomposes as,
\begin{equation}
S = S_{MF}(|\Delta_0|) + S_{\mathrm{ph}}(\theta),
\label{action}
\end{equation}
where $S_{MF}=S_0+S_{\mathrm{int}}(|\Delta_0|)$ is the mean-field action.  
In momentum space this reads,
\begin{align}
S_{MF}=\sum_{\mathbf{k},n}\ln\!\left[\det(-\mathbb{G}^{-1})_{\mathbf{k},n}\right]
+\beta\frac{1}{A}\sum_{\mathbf{k},\mathbf{p}}
\frac{|\Delta_0(\mathbf{k},T)||\Delta_0(\mathbf{p},T)|}{V_{eh}(\mathbf{k}-\mathbf{p})},
\label{SMF}
\end{align}
with $\mathbb{G}$ the Gor'kov single-particle propagator (details of the calculation are in Ref.~\cite{Pascucci2024thesis}). The phase action takes the form,
\begin{equation}
S_{\mathrm{ph}}=\frac{1}{2}\int_0^\beta\! d\tau\!\int d^{2}\bm{x}\,
\Bigl[J_0(T)\,(\nabla\theta)^{2}+\kappa_0(T)\,(\partial_\tau\theta)^{2}\Bigr]\ ,
\label{actionk}
\end{equation}
with $J_0(T)$ and $\kappa_0(T)$ defined in the main text.

The phase field $\theta$ can be decomposed into a longitudinal spin-wave component $\theta_{SW}$ and a
transverse static vortex component $\theta_V$,
\begin{equation}
\theta(\mathbf{x},\tau) = \theta_{SW}(\mathbf{x},\tau) + \theta_V(\mathbf{x})\ .
\label{decomp}
\end{equation}
$\theta_{SW}$ describes smooth phase fluctuations satisfying
$\nabla^{2}\theta_{SW}=0$, while $\theta_V$
encodes topological vortex configurations with quantized winding number,
$\nabla^{2}\theta_V=\sum_i 2\pi m_i\,\delta^{(2)}(\mathbf{x}-\mathbf{x}_i)$,
$m_i\in\mathbb{Z}$. The two contributions are orthogonal, so the phase action splits into  $S_{\mathrm{ph}}=S_{SW}+S_V$, with spin-wave contribution,
\begin{equation}
S_{SW}=\frac{1}{2}\int_0^\beta\! d\tau\!\int d^{2}\bm{x}\,
\Bigl[J_0(T)\,(\nabla\theta_{SW})^{2}+\kappa_0(T)\,(\partial_\tau\theta_{SW})^{2}\Bigr],
\label{SSW}
\end{equation}
and static vortex contribution
\begin{equation}
S_V=\frac{1}{2T}\int d^{2}\bm{x}\;
J_0(T)\,(\nabla\theta_V)^{2}\ .
\label{SV}
\end{equation}
Equation \eqref{SV} is formally equivalent to the classical $XY$ action governing the BKT transition \cite{Benfatto2004}. The vortex contribution $S_V$ is treated within the renormalization-group approach used in the main text to determine the
renormalized stiffness $J_{RG}(T)$ and the BKT transition temperature $T_{\mathrm{BKT}}^{\mathrm{RG}}$. It is not
discussed further here. 
\begin{figure}[h!]
\includegraphics[width=0.43\textwidth]{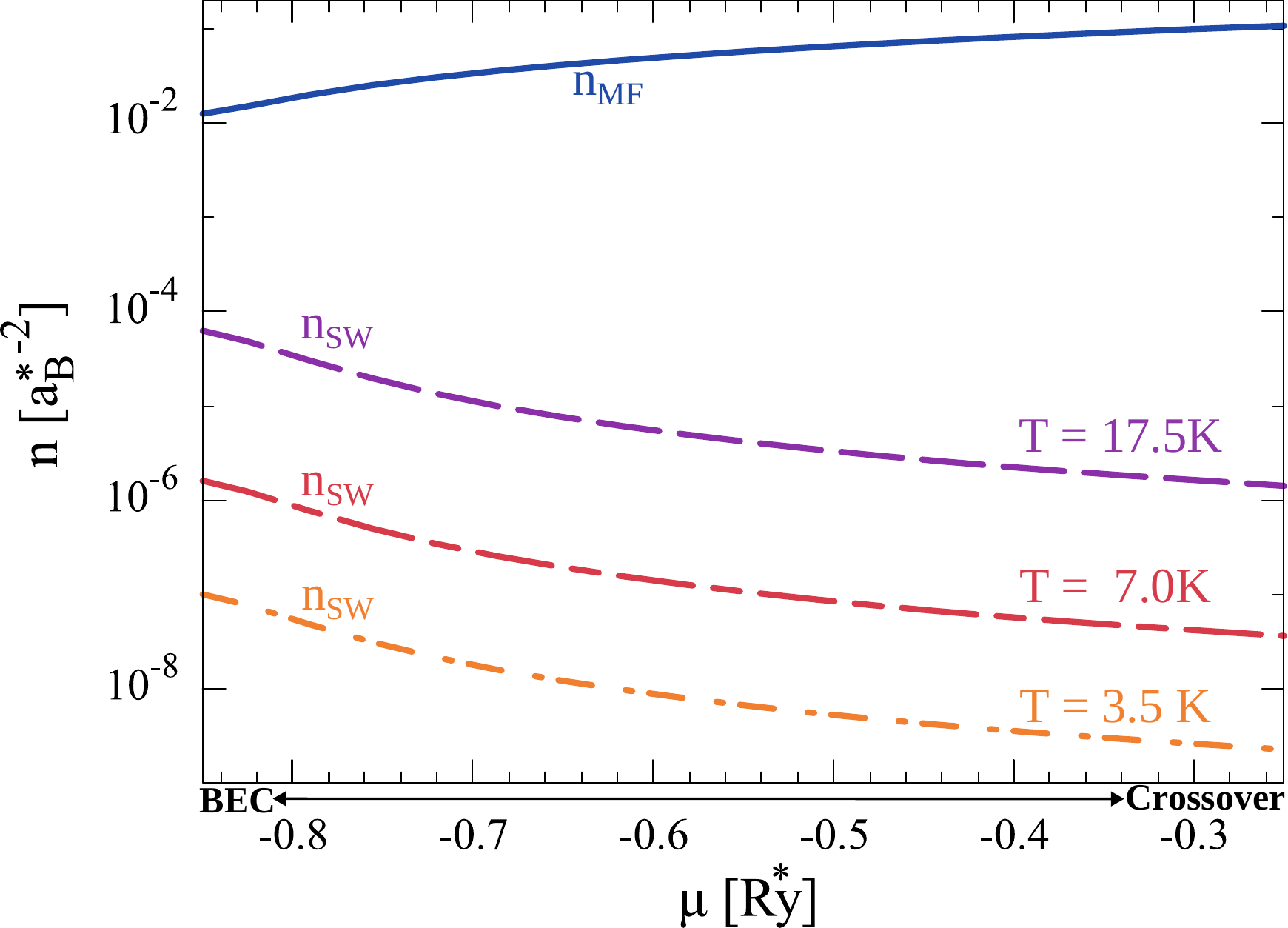}
\caption{Mean-field contribution $n_{MF}$ (solid line) and spin-wave
contribution $n_{SW}$ (dashed lines) to the density equation as a function of
the chemical potential $\mu$, evaluated using the mean-field solutions for
the gap and the chemical potential, in the BEC and crossover regimes. The
spin-wave contribution is shown for three representative temperatures , $T=3.5$, $7.0$, and $17.5\ \mathrm{K}$, of the
order of $T_{\mathrm{BKT}^{RG}}$ for the densities considered in this work.}
\label{phasefluct}
\end{figure}
The gap equation then follows from minimizing the mean-field grand potential $\Omega_{MF}=T S_{MF}$ with respect to
$\Delta_0$ (Eq.\ \eqref{gapeq}).

The density equation, which fixes the chemical potential, is obtained
from the full grand potential $\Omega=\Omega_{MF}+\Omega_{SW}$. Since the
spin-wave action $S_{SW}$ in Eq.~\eqref{SSW} is quadratic in the phase field,
the functional integration over $\theta_{SW}$ can be performed exactly,
yielding 
\begin{equation}
\Omega_{SW}=T\sum_{\mathbf{q}}\ln\!\left(1-e^{-\beta\omega(\mathbf{q})}\right),
\label{OmegaSW}
\end{equation}
with $\omega(\mathbf{q})=c_s q$ the collective-mode spectrum. From $n=-\frac{1}{A}\partial\Omega/\partial\mu_s$ one obtains $n=n_{MF}+n_{SW}$, with the mean-field contribution,
\begin{equation}
n_{MF}=\frac{g}{A}\sum_\mathbf{k}\frac{1}{2}\left(1-\frac{\epsilon_\mathbf{k}\tanh(E(\mathbf{k},T)/2T)}{E(\mathbf{k},T)}\right)
\end{equation}
coming from $-\frac{1}{A}\partial\Omega_{MF}/\partial\mu_s$, and the spin-wave
contribution,
\begin{equation}
n_{SW}=-\frac{1}{A}\frac{\partial\Omega_{SW}}{\partial\mu_s}\ .
\end{equation}
%

Figure~\ref{phasefluct} shows that, when evaluated using the mean-field solutions for the gap and chemical potential,
$n_{\mathrm{SW}}$ is several orders of magnitude smaller than $n_{\mathrm{MF}}$ over the temperature and density range
relevant to the superfluid phase. Up to temperatures of order $T_{\mathrm{BKT}}^{RG}$, spin-wave fluctuations, despite growing as the system approaches the low-density regime, provide only a negligible correction to the density equation. We can thus safely approximate $n\simeq n_{\mathrm{MF}}$ throughout this work.

\end{document}